\documentclass[twocolumn]{aastex63}

\usepackage{tablefootnote}

\received{xxx, xxx}
\revised{xxx, xxx}
\accepted{xxxx}
\submitjournal{AJ}

\shorttitle{The peculiar chemical pattern of the WASP-160 binary system}
\shortauthors{Jofr\'e et al.}

\begin{document}

\title{The peculiar chemical pattern of the WASP-160 binary system: signatures of planetary formation and evolution?}

\correspondingauthor{Emiliano Jofr\'e}
\email{emiliano@astro.unam.mx}

\author[0000-0002-8177-7633]{Emiliano Jofr\'e}

\affiliation{Universidad Nacional Aut\'onoma de M\'exico, Instituto de Astronom\'ia, AP 70-264, CDMX 04510, M\'exico} 


\affiliation{Universidad Nacional de C\'ordoba - Observatorio Astron\'{o}mico de C\'{o}rdoba, Laprida 854, X5000BGR, C\'ordoba, Argentina}

\affiliation{Consejo Nacional de Investigaciones Cient\'{i}ficas y T\'{e}cnicas (CONICET), Godoy Cruz 2290, CPC 1425FQB, CABA, Argentina}

\author{Romina Petrucci}
\affiliation{Universidad Nacional Aut\'onoma de M\'exico, Instituto de Astronom\'ia, AP 70-264, CDMX 04510, M\'exico} 

\affiliation{Universidad Nacional de C\'ordoba - Observatorio Astron\'{o}mico de C\'{o}rdoba, Laprida 854, X5000BGR, C\'ordoba, Argentina}

\affiliation{Consejo Nacional de Investigaciones Cient\'{i}ficas y T\'{e}cnicas (CONICET), Godoy Cruz 2290, CPC 1425FQB, CABA, Argentina}

\author{Yilen G\'omez Maqueo Chew}
\affiliation{Universidad Nacional Aut\'onoma de M\'exico, Instituto de Astronom\'ia, AP 70-264, CDMX 04510, M\'exico} 

\author{Ivan Ram\'irez}
\affiliation{Tacoma Community College, 6501 South 19th Street, Tacoma, WA 98466, USA}

\author{Carlos Saffe}
\affiliation{Instituto de Ciencias Astron\'omicas, de la Tierra y del Espacio, C.C 467, 5400, San Juan, Argentina}

\affiliation{Universidad Nacional de San Juan, Facultad de Ciencias Exactas, F\'isicas y Naturales, San Juan, Argentina}

\author{Eder Martioli}
\affiliation{Institut d'Astrophysique de Paris, UMR7095 CNRS, Universit\'e Pierre \& Marie Curie, 98 bis Boulevard Arago, 75014, Paris, France}

\affiliation{Laborat\'orio Nacional de Astrof\'isica, Rua Estados Unidos 154, Itajub\'a, MG, 37504-364, Brazil}

\author{Andrea P. Buccino}
\affiliation{Consejo Nacional de Investigaciones Cient\'{i}ficas y T\'{e}cnicas (CONICET), Godoy Cruz 2290, CPC 1425FQB, CABA, Argentina}

\affiliation{Universidad de Buenos Aires, Facultad de Ciencias Exactas y Naturales, Buenos Aires, Argentina}

\affiliation{CONICET - Universidad de Buenos Aires, Instituto de Astronom\'ia y F\'isica del Espacio (IAFE), Buenos Aires, Argentina}

\author{Martin Ma\v{s}ek}
\affiliation{FZU – Institute of Physics of the Czech Academy of Sciences, Na Slovance 1999/2, CZ-182 21 Praha, Czech Republic}

\author{Luciano Garc\'ia}
\affiliation{Universidad Nacional de C\'ordoba - Observatorio Astron\'{o}mico de C\'{o}rdoba, Laprida 854, X5000BGR, C\'ordoba, Argentina}

\author{Eliab Canul}
\affiliation{Universidad Nacional Aut\'onoma de M\'exico, Instituto de Astronom\'ia, AP 70-264, CDMX 04510, M\'exico} 

\author{Mercedes G\'omez}
\affiliation{Universidad Nacional de C\'ordoba - Observatorio Astron\'{o}mico de C\'{o}rdoba, Laprida 854, X5000BGR, C\'ordoba, Argentina}

\affiliation{Consejo Nacional de Investigaciones Cient\'{i}ficas y T\'{e}cnicas (CONICET), Godoy Cruz 2290, CPC 1425FQB, CABA, Argentina}




\begin{abstract}
Wide binary stars with similar components hosting planets provide a favorable opportunity for exploring the star-planet chemical connection. We perform a detailed characterization of the solar-type stars in the WASP-160 binary system. No planet has been reported yet around WASP-160A while WASP-160B is known to host a transiting Saturn-mass planet, WASP-160B b. For this planet, we also derive updated properties from both literature and new observations. Furthermore, using TESS photometry, we constrain the presence of transiting planets around WASP-160A and additional ones around WASP-160B. The stellar characterization includes, for the first time, the computation of high-precision differential atmospheric and chemical abundances of 25 elements based on high-quality Gemini-GRACES spectra. Our analysis reveals evidence of a correlation between the differential abundances and the condensation temperatures of the elements. In particular, we find both a small but significant deficit of volatiles and an enhancement of refractory elements in WASP-160B relative to WASP-160A. After WASP-94, this is the second stellar pair among the shortlist of planet-hosting binaries showing this kind of peculiar chemical pattern. Although we discuss several plausible planet formation and evolution scenarios for WASP-160A and B that could explain the observed chemical pattern, none of them can be conclusively accepted or rejected. Future high-precision photometric and spectroscopic follow-up, as  well  as  high-contrast  imaging  observations, of WASP-160A and B, might provide further constraints on the real origin of the detected chemical differences.

\end{abstract}



\section{Introduction} \label{sec:intro}
In the last decade, the analysis of high-precision differential abundances as a function of the condensation temperature of the elements (T$_{c}$) have revealed interesting chemical patterns that could be related to the planet formation processes and the subsequent planetary evolution. \citet[][hereafter M09]{Melendez2009} first found that the Sun is depleted in refractory elements (T$_{c}$ $\gtrsim$ 900 K) relative to volatiles (T$_{c}$ $<$ 900 K) when compared to a small sample of solar twins and that the abundance differences strongly correlate with T$_{c}$ \citep[see also][]{Ramirez2009}. M09 and \citet{Ramirez2009} suggested that this deficiency could indicate the formation of rocky bodies. In particular, they proposed that the missing refractories in the photosphere of the Sun were locked-up in rocky objects (e.g., asteroids, rocky planets) during the formation of the system. In recent years, other studies have derived similar results using considerably larger samples \citep[e.g.,][]{Bedell2018, Nibauer2020}. 

On the other hand, there is evidence of several stars showing T$_{c}$ trends with enhancement of refractory elements, relative to volatiles, when compared to field stars with similar atmospheric parameters \citep[e.g.,][]{Schuler2011, Melendez2017, Liu2020}. It has been proposed that this chemical anomaly might be produced by accretion events of already formed planets or planetesimals onto the central star \citep[e.g.,][]{Gonzalez1997, Pinsonneault2001, Sandquist2002}.

Alternatively, it has been argued that the observed T$_{c}$ correlations might be originated by other mechanisms not related to the presence of planets but instead rather be a consequence of other effects such as age and the Galactic birthplace \citep[e.g.,][]{Adibekyan2014, Nissen2015}, gas-dust segregation \citep{Gaidos2015} or radiative dust-cleansing \citep{Onehag2011, Onehag2014}. 

One way to mitigate or even remove these effects is by studying binary systems. Observations and numerical simulations support the idea that the components of these systems form coevally from the same molecular cloud and share the initial chemical composition \citep[e.g.,][]{Desidera2006, Kratter2011, Reipurth2012, King2012, Vogt2012, Alves2019, Andrews2019, Hawkins2020, Nelson2021}. In particular, the analysis of wide binary systems with similar components, where at least one of them hosts a planet, are interesting targets to explore the impact of planet formation and evolution on the stellar composition regardless of environmental or Galactic chemical evolution effects. 

To date, only five wide binaries with planets around one component have been previously studied in the literature: 16 Cyg \citep[e.g.,][]{Ramirez2011, Tucci2014, Tucci2019}; HAT-P-1 \citep{Liu2014}; HD 80606 \citep{Saffe2015}; HAT-P-4 \citep{Saffe2017}; and HD 106515 \citep{Saffe2019}. The other analyzed binaries host planets around both stars: HD 20781-82 \citep{Mack2014}; XO-2 \citep[e.g.,][]{Ramirez2015, Biazzo2015}; WASP-94 \citep{Teske2016a}; and HD 133131 \citep{Teske2016b}.

Here we expand the small sample of well-studied planet-hosting binary stars by investigating, for the first time, the chemical composition of the WASP-160 binary system. A Saturn-mass planet ($M_{\mathrm{p}}$ $\sim$ 0.28 $M_{\mathrm{J}}$, where $M_{\mathrm{J}}$ is the mass of Jupiter) was recently discovered to transit WASP-160B at 0.045 AU \citep[][hereafter LE19]{Lendl2019} via the WASP-South transiting planet detection program \citep{Hellier2011}. For WASP-160A, LE19 found no evidence of large amplitude radial velocity (RV) variations nor signs of any transiting planet. They also notice that WASP-160A (V = 12.7 mag) and WASP-160B (V = 13.1 mag) have consistent systemic RVs, proper motions and parallaxes from \textit{Gaia} \citep{Gaia2018}, indicating that the components are likely physically bound (see also Section \ref{sec.abundances}). Additionally, these authors reported that both stars appear to have similar masses and early K spectral types with a projected separation of $\sim$28.5 arcsec which translates into a physical projected distance of 8060 AU. 

In Section \ref{sec.data}, we describe our spectroscopic and photometric observations. In Section \ref{sec.stellar}, we present the stellar characterization analysis that includes the computation of high-precision elemental abundances for 25 elements, whilst in Section \ref{sec.differences.abundances}, we investigate the chemical differences between the components of the WASP-160 system. In Section \ref{sec.planetary}, we analyze our precise stellar parameters in combination with RV and photometric data to refine the planetary parameters of WASP-160B b. Also, from TESS data, we search for transiting planets around WASP-160A and for additional ones orbiting WASP-160B. In Section \ref{sec.discussion}, we discuss some scenarios that could explain the observed chemical pattern and, finally, in Section \ref{sec.conclussions} we present our main conclusions.

\begin{figure*}[ht!]
\centering
\includegraphics[width=.45\textwidth]{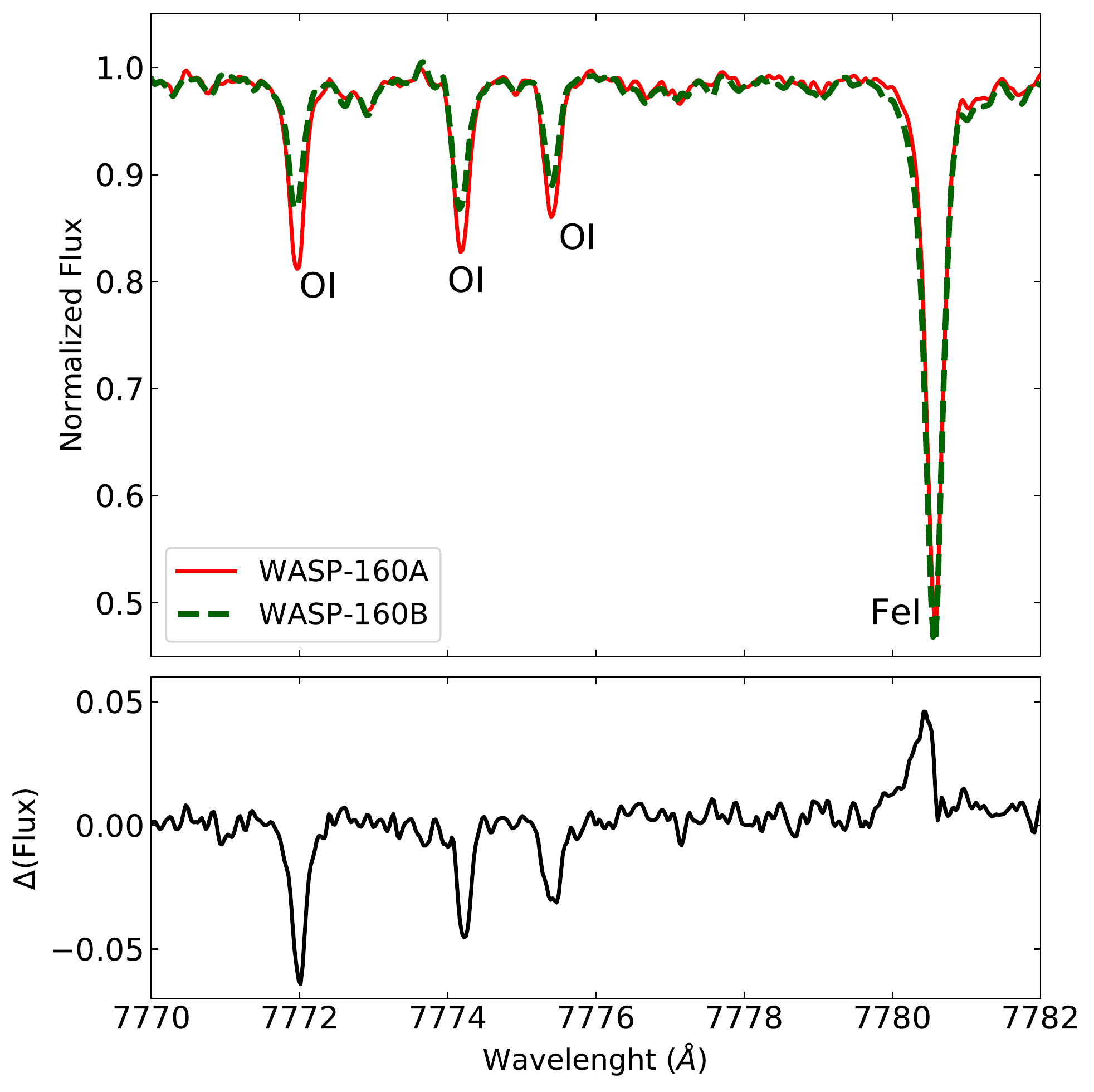}
\includegraphics[width=.45\textwidth]{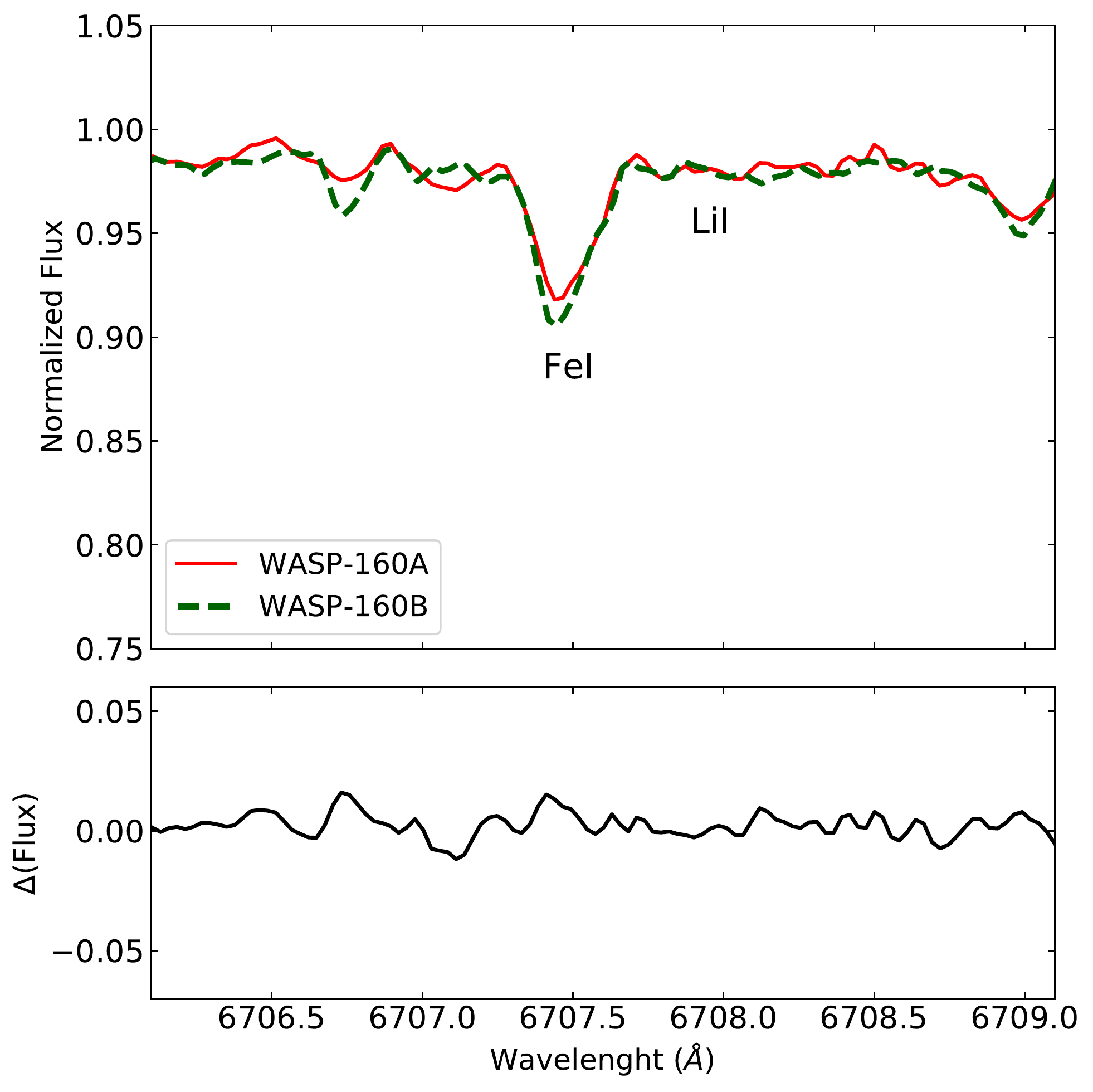}
\caption{Top panels: spectra of WASP-160A (red solid) and WASP-160B (green dashed) around the oxygen triplet region (left) and the 6708 {\AA} lithium line (right). Bottom panels: normalized flux difference between the WASP-160A and B. \label{fig.spectra}}
\end{figure*}

\section{Observations and literature data}\label{sec.data}
In this section, we present the newly acquired observations and describe their data reduction. We also summarize all the data from the literature utilized in the global analysis of the WASP-160B system, including the following photometric and spectroscopic observations taken at ESO La Silla Observatory in Chile, already published in LE19:

\begin{itemize}
    \item One partial transit observed in the I+z'-band the night of 16 Dec 2014 with the 0.6-m TRAPPIST telescope.
   \item One complete and one partial transit observed the nights of 26 Dec 2015 and 02 Jan 2017, respectively, in the r'-band with the EulerCam at the 1.2-m Euler-Swiss telescope.
   \item 30 radial velocities (RV) obtained with the CORALIE echelle spectrograph at the 1.2-m Euler-Swiss telescope.
\end{itemize}

\subsection{High-resolution Spectroscopy} \label{sec.spec.observations}
We observed WASP-160A and WASP-160B on 2019 November 13 UT with the Gemini Remote Access to CFHT ESPaDOnS Spectrograph \citep[GRACES;][]{Chene2014} at the 8.1-m Gemini North telescope. Observations were carried out in the queue mode (GN-2019B-Q-116, PI: E. Jofr\'e) in the one-fiber mode (object only), which achieves a resolving power of R$\sim$ 67,500 between 400 and 1,050 nm. We obtained consecutive exposures of 4 $\times$ 1060 s for WASP-160A and 4 $\times$ 1469 s for WASP-160B. 

These observations along with a series of calibrations including 6 $\times$ ThAr arc lamp, 10 $\times$ bias, and 7 $\times$ flat-field exposures were used as input in the OPERA (Open source Pipeline for ESPaDOnS Reduction and Analysis) code\footnote{OPERA is available at \url{http://wiki.lna.br/wiki/espectro}.} software \citep{Martioli2012} to obtain the reduced data. The reduction details can be found in \citet{Jofre2020}. Briefly, the reduction includes optimal extraction of the orders, wavelength calibration, and normalization of the spectra.  

Using the IRAF\footnote{IRAF is distributed by the National Optical Astronomy Observatories, which are operated by the Association of Universities for Research in Astronomy, Inc., under cooperative agreement with the National Science Foundation.} task \texttt{scombine}, the individual exposures of each target were co-added to obtain the final spectra which present a signal-to-noise ratio per resolution element of S/N $\sim$ 350 and S/N $\sim$ 320 around 6000 \AA, for WASP-160A and WASP-160B, respectively. We derived absolute radial velocities (RVs) by cross-correlating our program stars with standard stars using the IRAF task \texttt{fxcor}, obtaining VR = $-$6.28 $\pm$ 0.03 km s$^{-1}$ for the primary and VR = $-$6.30 $\pm$ 0.06 km s$^{-1}$ for the secondary. These values are in excellent agreement with the absolute RVs from \textit{Gaia} Data Release 2  \citep[DR2;][]{Gaia2018} and those from LE19. The combined spectra were corrected for the RVs shifts with the IRAF task \texttt{dopcor}. Small portions of the WASP-160A and WASP-160B final spectra are shown in Figure \ref{fig.spectra}. 

For the spectroscopic analysis below, we also employed a solar spectrum (reflected sunlight from the Moon) obtained with the same GRACES setup (SNR $\sim$ 410 at 6000 \AA) but observed on a different date. Although this is not ideal, given that the components of the binary system are not solar twins (see Section \ref{sec.atmospheric.parameters}), the highest precision in the chemical abundances is obtained from the differential analysis between WASP-160A and B as we show in Section \ref{sec.atmospheric.parameters}.


\subsection{Time-series Photometric Observations}

In order to refine the planetary parameters of WASP-160B b, we obtained six new full transits: 
one of them was observed by our team with the FRAM telescope and the other five were observed by the Transiting Exoplanet Survey Satellite \citep[TESS;][]{Ricker2015}.

\subsubsection{FRAM}{\label{fram}}

One full transit of WASP-160B b was observed in the R-filter the night of 2020 February 19 with the 0.3-m FRAM (F/(Ph)otometric Robotic Atmospheric Monitor) telescope, which is part of the Pierre Auger Observatory in Argentina. FRAM is a 0.3-m Orion Optimised Dall-Kirkham, f/6.8 with Moravian Instruments CCD MII G4-16000, 4096$\times$4096 pixels, and a field of view of 62\arcmin$\times$62\arcmin, and photometric BVRI filters \citep{Janecek2019}. At the end of the night, we took dark and sky-flat frames. A master dark was subtracted from the 124 science images and then divided by a master flat created as the median combined dark-corrected individual sky-flats using standard IRAF routines. As in \citet{Petrucci2020}, we used the FOTOMCAp code \citep{Petrucci2016} to measure precise instrumental magnitudes and then selected three non-variable stars in the field as comparisons to build up the differential light curve with the lowest dispersion in magnitudes. 

In order to choose the best detrending model to the photometric data,   we first subtracted a JKTEBOP \citep{Southworth2004} transit model from the resulting observed light curve. This transit model was computed by fixing the inclination, sum and ratio of the relative planetary and stellar radii, and limb-darkening coefficients to the values determined by LE19. Airmass, background counts, FWHM, spatial shifts (in X and Y), peak-value, and seeing were measured for each timestamp. We created different detrending models as linear combinations of these variables and fitted the residuals. The final adopted light curve of WASP-160B (Figure \ref{fig.groundLCRV}) was the light curve corrected by the model that minimized the chi-square of the residuals, in this case, the one that combined all the 7 variables.


\subsubsection{TESS}{\label{tess}}

\begin{figure}
\centering
\includegraphics[width=.45\textwidth]{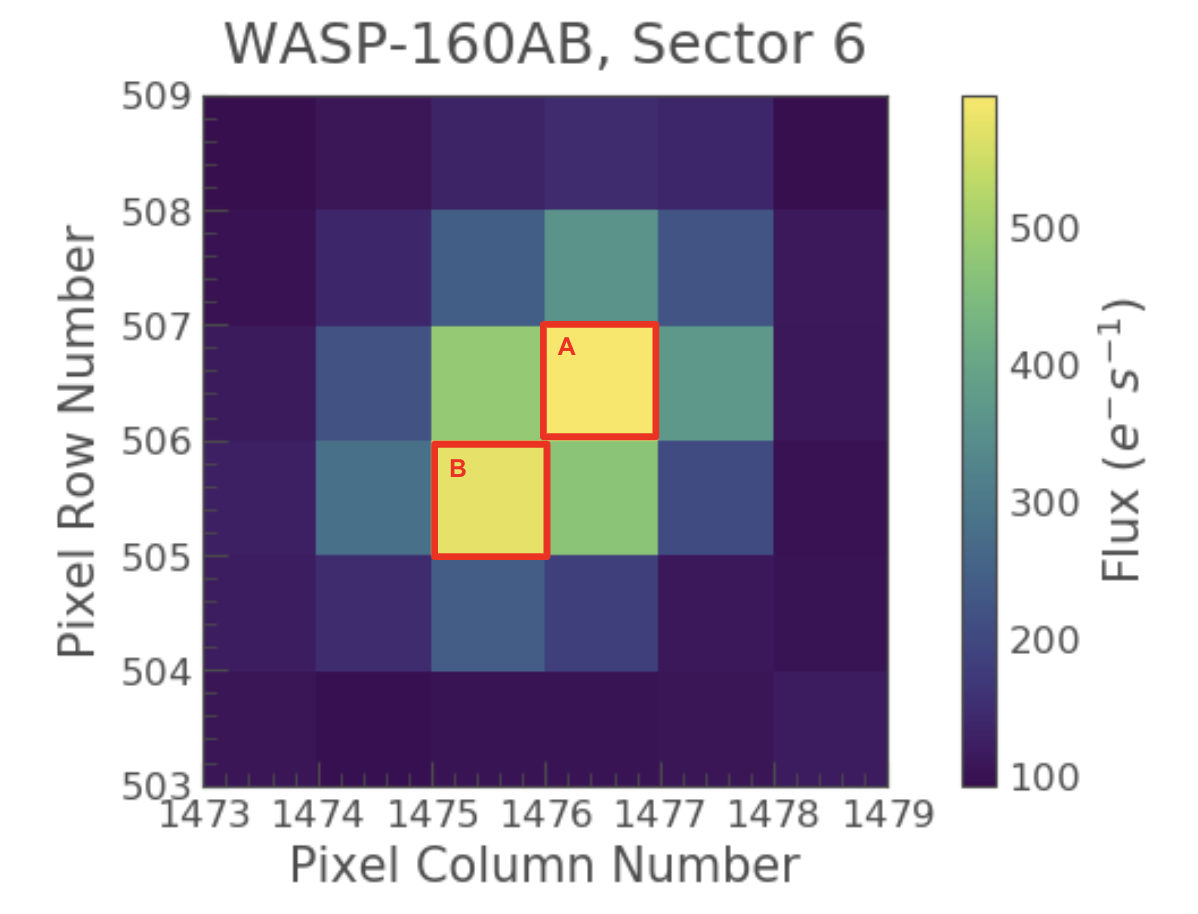}
\caption{6$\times$6 pixel TESS FFI centered on WASP-160AB. Light and dark colors show pixels with high and low flux values, respectively. Red squares and letters indicate the pixels selected to perform the aperture photometry of both stars (see the text for details).\label{fig.w160AB.tess}}
\end{figure}

WASP-160AB was observed by TESS with 30-min-cadence in sector 6 (2018 December 11 to 2019 January 07). These data, publicly available in the form of full frame images (FFI), were analyzed with the tools provided  by the \texttt{Lightkurve} Python package \citep{Lightkurve}. We performed single aperture photometry on the 987 available FFI, only considering the square pixel with the highest flux value centered on the planet host-star, WASP-160B (Figure \ref{fig.w160AB.tess}). To remove the systematics owing to scattered light and spacecraft instrument/motion noise\footnote{The Data Release Notes of the sector do not report any additional systematics that we had to take into account in the detrending process.}, we applied a Savitzky-Golay filter \citep{SavGol} provided by \texttt{Lightkurve} that, basically, smooths the light curve by fitting a polynomial at consecutive intervals. In Figure \ref{fig.tessLC}, we present the final phase folded-transit light curve of WASP-160B considering the five individual transits found in the sector 6 of TESS.

Given that the size of the TESS pixel is 21 arcsec and the separation between the components of the system is $\sim$28 arcsec, the selected aperture includes the flux contribution of both stars\footnote{We used the \texttt{tpfplotter} code \citep{Aller2020} to visually inspected if, besides WASP-160A, there was another nearby \textit{Gaia} DR2 source that might be contaminating the selected aperture or even be included in it. We found the star TIC 32949758 located at $\sim$23 arcsec from WASP-160B. However, it is at least 5 times fainter in the \textit{Gaia} band and, then, produces negligible light contamination into the selected photometric aperture.}. This light contamination from a blended stellar source makes the measured transit depth shallower than it truly is. In order to avoid a wrong determination of the planetary radius, we fitted a  parameter that takes into account this effect during the light curve modeling as is explained in Section {\ref{exofastv2}}. This allowed us to correctly estimate the radius of the planet from the TESS data.





\section{Stellar Characterization} \label{sec.stellar}

\subsection{Spectroscopic Fundamental Parameters} \label{sec.atmospheric.parameters}
The atmospheric parameters T$_{\mathrm{eff}}$ (effective temperature), $\log g$ (surface gravity), [Fe/H] (abundance of iron), and $v_{micro}$ (microturbulent velocity) were derived from equivalent widths (EWs) of iron lines  by imposing differential excitation and ionization equilibrium, as described in previous works \citep[e.g.,][]{Ramirez2015, Ramirez2019, Saffe2015, Saffe2017}. Briefly, in this method, the T$_{\mathrm{eff}}$ and $v_{micro}$ are iteratively modified until the correlation of differential iron abundance (Fe \textsc{i} lines only) with excitation potential (EP=$\chi$) and reduced equivalent width (REW = $\log \mathrm{EW}/\lambda$), respectively, are minimized. Simultaneously, $\log g$ is modified until the differential abundance derived from Fe \textsc{i} lines approaches to the one computed from Fe \textsc{ii} lines. To perform this process automatically, we employed the \texttt{Qoyllur-Quipu} (or \texttt{q$^{2}$}) Python package\footnote{The code is available at \url{https://github.com/astroChasqui/q2}} \citep{Ramirez2014}, which uses the \texttt{MOOG} code \citep{Sneden1973} to translate the EW measurements into abundances via the curve-of-growth method. In addition, \texttt{q$^{2}$} uses linearly interpolated one-dimensional local thermodynamic equilibrium (LTE) Kurucz ODFNEW model atmospheres \citep{Castelli2003}. 

The iron line list, as well as the atomic parameters (EP and oscillator strengths, $\log gf$), were compiled from \citet{Ramirez2015} and the EWs were manually measured by fitting Gaussian profiles using the \texttt{splot} task in IRAF. To make a precise and consistent measurement of the EWs in the spectra of WASP-160A, WASP-160B and the Sun, for each line, we carefully selected the same continuum region line by overplotting the spectra of the stars. 

\begin{figure}[ht!]
\centering
\includegraphics[width=.45\textwidth]{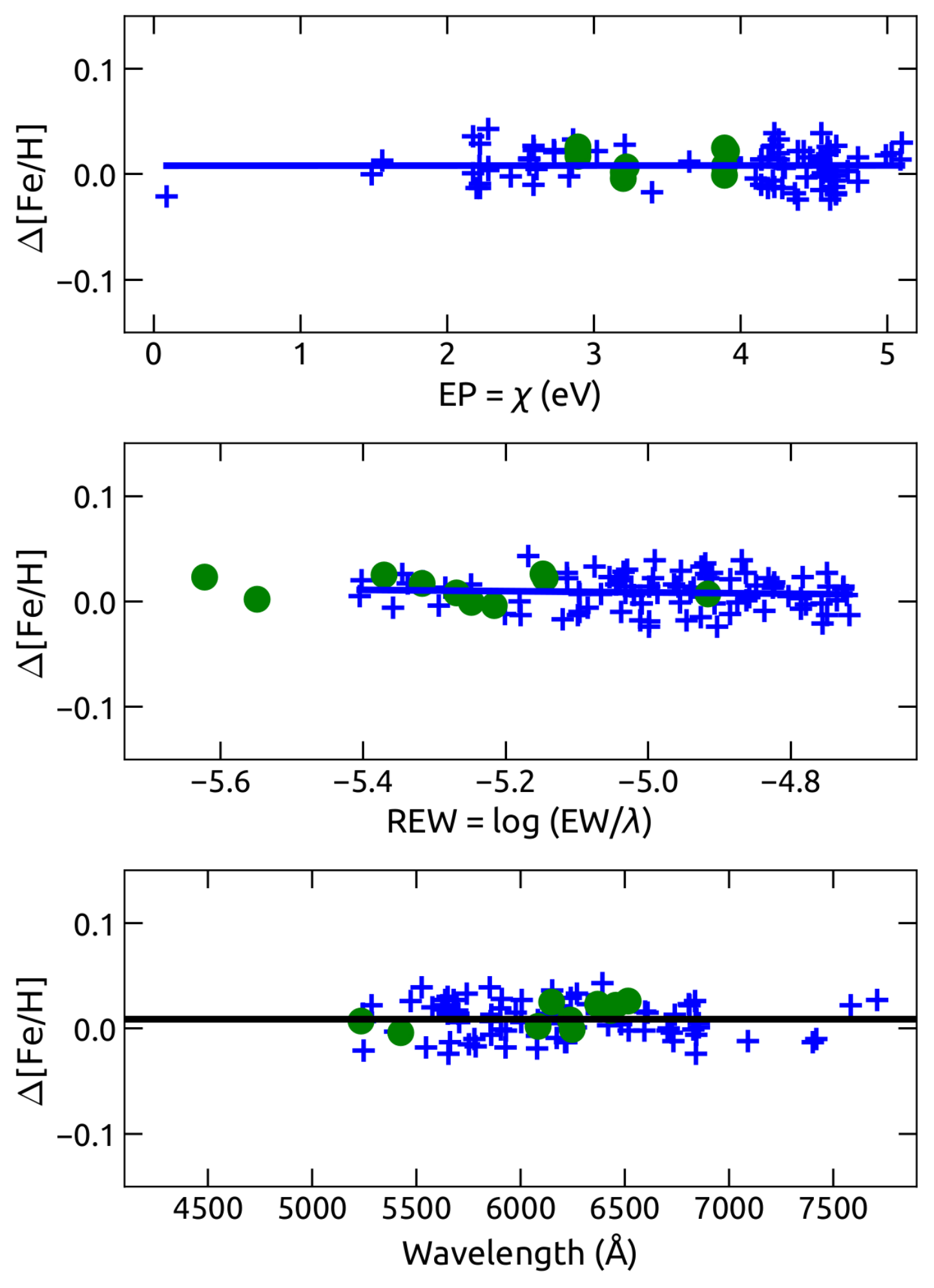}
\caption{Line-by-line iron abundance of WASP-160B, computed relative to WASP-160A, as a function of excitation potential (upper panel), reduced EW (middle panel), and wavelength (bottom panel). Blue crosses (green circles) correspond to Fe \textsc{i} (Fe \textsc{ii}) lines. In the top and middle panel, solid lines are linear fits to the Fe \textsc{i} data, while the solid black line in the bottom panels indicates the average iron abundance from all lines. \label{fig.equilibrium}}
\end{figure}

We first performed the line-by-line differential analysis of both components of the stellar pair using our solar spectrum as reference, adopting as solar parameters T$_{\mathrm{eff}}$ = 5777 K, $\log g$ = 4.44 dex, [Fe/H] = 0 dex, and $v_{micro}$ = 1.0 km s$^{\mathrm{-1}}$. As initial values we employed those obtained by LE19, which are included in the first block of Table \ref{table.spectroscopic.parameters}. The results of this computation are listed in the second block of Table \ref{table.spectroscopic.parameters}. We note that only the [Fe/H] value for WASP-160A and the T$_{\mathrm{eff}}$ of WASP-160B computed by LE19 are consistent, within the error bars, with our results. The rest of the parameters agree only within the 2$\sigma$ range. In particular, the $\log g$ and [Fe/H] values are systematically larger than ours. The discrepancies are likely related to the use of different techniques (i.e., differential vs. absolute), models of stellar atmospheres, line-lists and atomic data, and the use of spectra of different quality \citep[e.g.,][]{Bedell2014, Hinkel2016, Jofre2017, Jofre2020}.

From our first differential results (second block of Table \ref{table.spectroscopic.parameters}), we observe that both components present very similar $\log g$ and [Fe/H] values ($\Delta\log g \sim 0.02$ dex and $\Delta$[Fe/H] $\sim$ 0.02 dex). The WASP-160 stellar components present a difference in effective temperature of $\Delta$T$_{\mathrm{eff}}$ $\sim$ 200 K, which after the system HD 20781/82 with $\Delta$T$_{\mathrm{eff}}$ $\sim$ 465 K \citep{Mack2014}, is the stellar pair with the largest difference in T$_{\mathrm{eff}}$ among the sample of binaries hosting planetary systems with high-precision chemical abundances \citep[e.g.,][]{Ramirez2019}. Even with the large difference in T$_{\mathrm{eff}}$, the components of the WASP-160 binary system can still be considered as twin stars according to the criteria of \citet[][$\Delta$T$_{\mathrm{eff}}$ $\lesssim$ 300 K and $\Delta$$\log g$ $\lesssim$ 0.2 dex]{Nagar2020}. The measured $\Delta$T$_{\mathrm{eff}}$ might account for the slightly larger errors in the derived fundamental parameters, especially the abundances of iron (and other elements), in comparison with previous differential analysis of stellar twins \citep{Ramirez2015, Teske2016a, Saffe2017}. In Section \ref{effect.Teff}, we discuss the impact of the derived T$_{\mathrm{eff}}$ on the abundance differences between the binary components.

\begin{table*}
\caption{Atmospheric parameters of WASP-160A and WASP-160B}
\label{table.spectroscopic.parameters}
\centering
\begin{tabular}{l c c c c}
\hline\hline
Star	&	$T_{\mathrm{eff}}$			&	$\log g$ 			&	$\mathrm{[Fe/H]}$ 			&	$v_{\mathrm{micro}}$ 			\\
	&	(K)			&	(cgs)			&	(dex)			&	(km $s^{-1}$)			\\
\hline																	
\multicolumn{5}{c}{\citet{Lendl2019}} \\																	
\hline																	
WASP-160A	&	5300	$\pm$	100	&	4.50	$\pm$	0.20	&	0.19	$\pm$	0.09	&	$-$			\\
WASP-160B	&	5300	$\pm$	100	&	4.60	$\pm$	0.10	&	0.27	$\pm$	0.10	&	$-$			\\
$\Delta$(B$-$A)	&	0	$\pm$	141	&	0.10	$\pm$	0.22	&	0.08	$\pm$	0.13	&	$-$			\\
\hline																	
\multicolumn{5}{c}{This work, solar reference (Section \ref{sec.atmospheric.parameters})}\\																	
\hline																	
WASP-160A &	5424	$\pm$	16	&	4.42	$\pm$	0.05	&	0.139	$\pm$	0.017	&	0.91 $\pm$		0.06	\\
WASP-160B	&	5218	$\pm$	18	&	4.44	$\pm$	0.05	&	0.149	$\pm$	0.020	&	0.76 $\pm$		0.09	\\
$\Delta$(B$-$A)	&	$-$206	$\pm$	24	&	0.02	$\pm$	0.07	&	0.01	$\pm$	0.026	&	$-$0.15	$\pm$	0.11	\\
\hline																	
\multicolumn{5}{c}{This work, WASP-160A reference (Section \ref{sec.atmospheric.parameters})}\\																	
\hline																	
WASP-160A	&	5424	$\pm$	16	& 	4.42	$\pm$	0.05	&	0.139	$\pm$	0.017	&	0.91	$\pm$	0.06	\\
WASP-160B	&	5215	$\pm$	4	&	4.47	$\pm$	0.02	&	0.151	$\pm$	0.004	&	0.75	$\pm$	0.03	\\
$\Delta$(B$-$A)	&	$-$209	$\pm$	16	&	0.05	$\pm$	0.05	&	0.012	$\pm$	0.017	&	$-$0.16	$\pm$	0.07	\\
\hline																																		
\end{tabular}

\end{table*}

\begin{figure*}[ht!]
\centering

\includegraphics[width=.49\textwidth, trim = {0.8cm 0.4cm 0.0cm 0.0cm}]{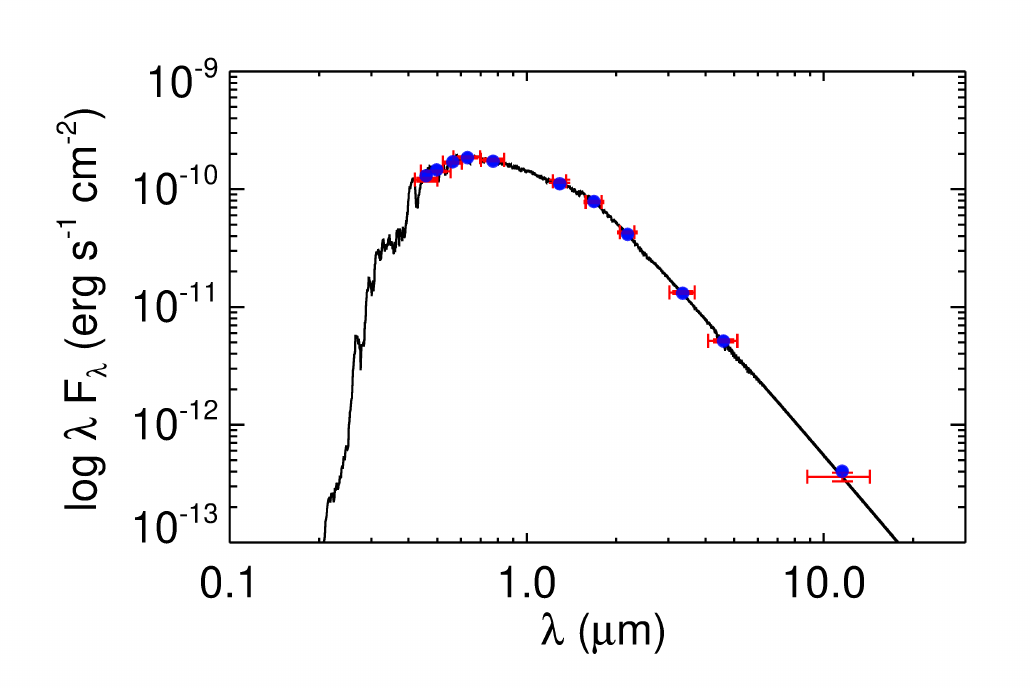}
\includegraphics[width=.49\textwidth, trim = {0.8cm 0.4cm 0.0cm 0.0cm}]{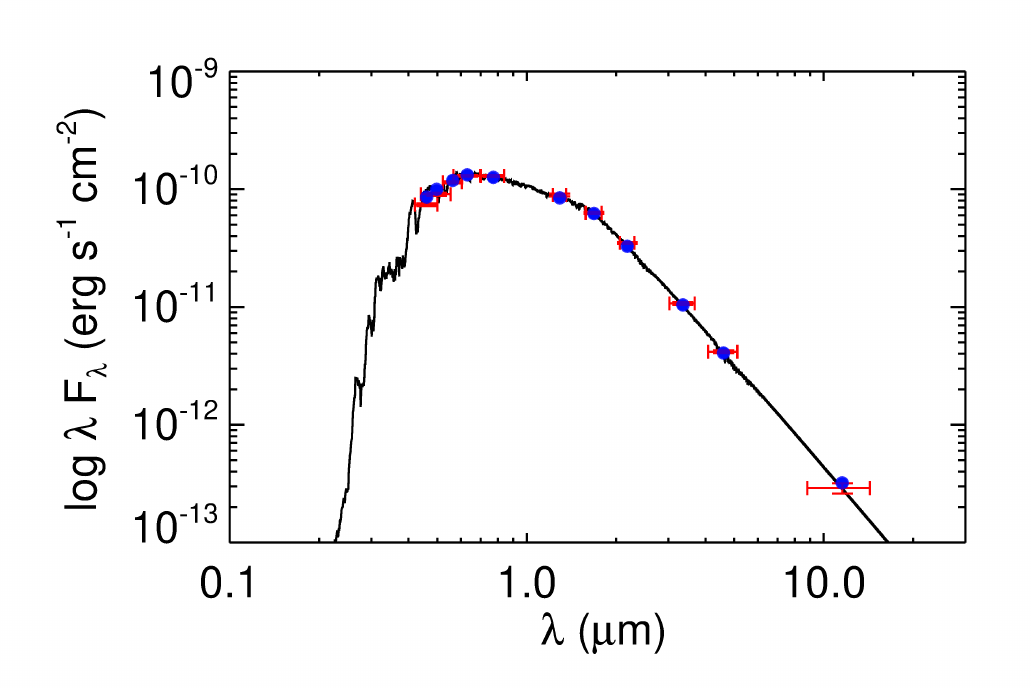}
\caption{Spectral energy distributions of WASP-160A (left) and WASP-160B (right). The red points represent the measurements of the fluxes in the WASP-160 stars in the optical, NIR, and mid-IR passbands listed in Table \ref{table.all.stellar.parameters}. The vertical bars are the $1\sigma$ photometric uncertainties and the horizontal error bars indicate the width of the photometric bandpasses.  Blue symbols are the model fluxes from the best-fit NextGen atmosphere model (black).
 \label{fig.SED}}
\end{figure*}

On the other hand, it can also be noticed that WASP-160A and WASP-160B are significantly different from the Sun in terms of T$_{\mathrm{eff}}$ (i.e, non-solar twins) and therefore it would be expected to obtain more precise parameters by comparing the two stars differentially to each other. Hence, we repeated the process to measure the stellar parameters of WASP-160B but considering the star WASP-160A as reference by adopting its solar-reference parameters derived above as fixed. We choose WASP-160A as the reference star since its solar-relative atmospheric parameters (especially the T$_{\mathrm{eff}}$) are more similar to those of the Sun and therefore expected to be more reliable than those of the B component \citep[see][]{Teske2016a}. The final result (WASP-160B $-$ WASP-160A) is shown in Figure \ref{fig.equilibrium} and the derived parameters with their errors are listed in the last block of Table \ref{table.spectroscopic.parameters}. As expected, the atmospheric parameters of WASP-160B are almost the same than those computed using the Sun as reference, but the errors are considerably smaller. These errors correspond to intrinsic uncertainties of the method and are estimated as in \citet{Epstein2010} and \citet{Bensby2014}. Both the results derived comparing the WASP-160 stars to the Sun and those obtained through a strict differential analysis of WASP-160B relative to WASP-160A reveal that the B component is only marginally iron-enhanced relative to A by $\sim$+0.01 dex.

 \begin{table*}
  \small
      \caption{Stellar properties of WASP-160A and WASP-160B}
         \label{table.all.stellar.parameters}
     \centering
         \begin{tabular}{l c c c}
            \hline\hline
            
Parameter	&	WASP-160A			&	WASP-160B			&	Source	\\
\hline											
\multicolumn{4}{c}{Astrometry and kinematics} \\											
\hline											
Proper motion in RA, $\mu_{\alpha}$ (mas yr$^{-1}$)	&	26.854	$\pm$	0.034	&	26.975	$\pm$	0.030	&	\textit{Gaia} DR2	\\
Proper motion in DEC, $\mu_{\delta}$ (mas yr$^{-1}$)	&	$-$34.821	$\pm$	0.042	&	$-$34.830	$\pm$	0.037	&	\textit{Gaia} DR2	\\
Parallax, $\pi$ (mas)	&	3.38	$\pm$	0.06	&	3.36	$\pm$	0.06	&	\textit{Gaia} DR2	\\
Barycentric radial velocity, RV (km $s^{-1}$)	&	$-$6.28	$\pm$	0.03	&	$-$6.30	$\pm$	0.06	&	This work (Sec. \ref{sec.spec.observations})	\\
Space velocity component, U (km $s^{-1}$)	&	$-$6.57	$\pm$	0.02	&	$-$6.59	$\pm$	0.03	&	This work (Sec. \ref{sec.abundances})	\\
Space velocity component, V (km $s^{-1}$)	&	2.96	$\pm$	0.02	&	2.97	$\pm$	0.04	&	This work (Sec. \ref{sec.abundances})	\\
Space velocity component, W (km $s^{-1}$)	&	1.05	$\pm$	0.01	&	1.06	$\pm$	0.03	&	This work (Sec. \ref{sec.abundances})	\\
Galaxy population membership	&	Thin  disk			&	Thin  disk			&	This work (Sec. \ref{sec.abundances})	\\
\hline											
\multicolumn{4}{c}{Photometry} \\											
\hline											
B (mag)	& 	13.452	$\pm$	0.040	& 	13.983	$\pm$	0.030	& 	APASS	\\
V (mag)	& 	12.677	$\pm$	0.020	& 	13.094	$\pm$	0.020	& 	APASS	\\
$g'$ (mag) 	& 	13.014	$\pm$	0.020	& 	13.500	$\pm$	0.030	& 	APASS	\\
$r'$ (mag)	& 	12.426	$\pm$	0.020	& 	12.820	$\pm$	0.020	& 	APASS	\\
$i'$ (mag)	& 	12.253	$\pm$	0.030	& 	12.595	$\pm$	0.020	& 	APASS	\\
J (mag)	& 	11.300	$\pm$	0.030	& 	11.591	$\pm$	0.030	& 	2MASS	\\
H (mag)	& 	10.937	$\pm$	0.020	& 	11.172	$\pm$	0.020	& 	2MASS	\\
Ks (mag) 	& 	10.831	$\pm$	0.020	& 	11.055	$\pm$	0.020	& 	WISE	\\
$W_1$  (mag)	& 	10.789	$\pm$	0.030	& 	11.021	$\pm$	0.030	& 	WISE	\\
$W_2$ (mag)	& 	10.834	$\pm$	0.030	& 	11.067	$\pm$	0.030	& 	WISE	\\
$W_3$ (mag)	& 	10.798	$\pm$	0.092	& 	11.038	$\pm$	0.110	& 	WISE	\\
V-band extinction, A$_{V}$ (mag) 	&	0.0308	$\pm$	0.015	&	0.0308	$\pm$	0.0015	&	\citet{Schlafly2011}	\\
\hline											
\multicolumn{4}{c}{Adopted atmospheric and physical  parameters} \\											
\hline											
Effective temperature, $T_{\mathrm{eff}}$ (K)	&	5424	$\pm$	16	&	5215	$\pm$	4	&	This work (Sec. \ref{sec.atmospheric.parameters})	\\
Surface gravity, $\log g$ (cgs) 	&	4.42	$\pm$	0.05	&	4.47	$\pm$	0.02	&	This work (Sec. \ref{sec.atmospheric.parameters})	\\
Metallicity, $\mathrm{([Fe/H])}$ 	&	0.139	$\pm$	0.017	&	0.151	$\pm$	0.004	&	This work (Sec. \ref{sec.atmospheric.parameters})	\\
Microturbulent velocity, $v_{\mathrm{micro}}$ (km $s^{-1}$)	&	0.91	$\pm$	0.06	&	0.75	$\pm$	0.03	&	This work (Sec. \ref{sec.atmospheric.parameters})	\\
Macroturbulent velocity, $v_{\mathrm{macro}}$ (km $s^{-1}$)	&	4.51	$\pm$	0.032	&	4.83	$\pm$	0.025	&	This work (Sec. \ref{sec.atmospheric.parameters})	\\
Rotational velocity, $v\sin i$ (km $s^{-1}$)	&	3.96	$\pm$	0.30	&	3.61	$\pm$	0.48	&	This work (Sec. \ref{sec.atmospheric.parameters})	\\
Activity index $\log R'_{HK}$	&	$-$4.714	$\pm$	0.0434	&	$-$4.766	$\pm$	0.0434	&	This work (Sec. \ref{sec.activity})	\\
Mass, $M_{\mathrm{\star}}$ ($M_{\mathrm{\odot}}$)	&	0.926	$^{+0.024}_{-0.022}$		&	0.879	$^{+0.022}_{-0.020}$		&	This work (Sec. \ref{sec.other.parameters})	\\
Radius, $R_{\mathrm{\star}}$ ($R_{\mathrm{\odot}}$)	&	0.950	$^{+0.018}_{-0.019}$		&	0.872	$^{+0.013}_{-0.012}$		&	This work (Sec. \ref{sec.other.parameters})	\\
Age, $\tau$ (Gyr)	&	8.0	$^{+2.5}_{-2.6}$		&	8.2	$^{+2.8}_{-3.0}$		&	This work (Sec. \ref{sec.other.parameters})	\\
Luminosity, $L_{\mathrm{\star}}$ ($L_{\mathrm{\odot}}$)	&	0.69	$\pm$	0.026	&	0.502	$^{+0.015}_{-0.014}$		&	This work (Sec. \ref{sec.other.parameters})	\\
Density, $\rho_{\mathrm{\star}}$ (cgs)	&	1.522	$^{+0.11}_{-0.099}$		&	1.869	$^{+0.095}_{-0.088}$		&	This work (Sec. \ref{sec.other.parameters})	\\

            \hline
         \end{tabular} 
     
        \end{table*}

\begin{figure}[ht!]
\centering
\includegraphics[width=.49\textwidth]{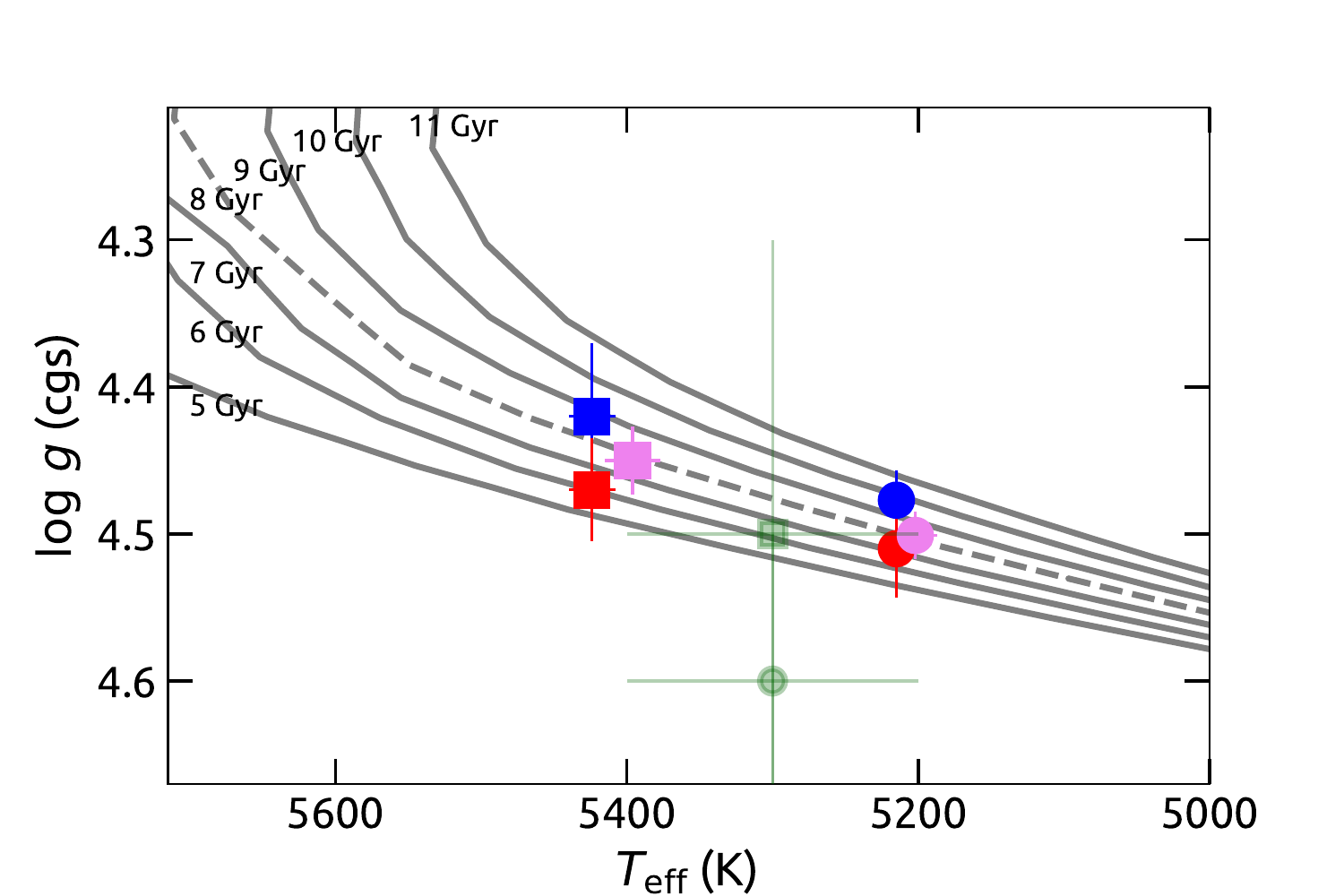}
\caption{Locations of WASP-160A (squares) and WASP-160B (circles) in the T$_{\mathrm{eff}}$--$\log g$ plane using different sets of parameters. The blue symbols show the purely differential spectroscopic parameters derived in this work. The pink symbols represent T$_{\mathrm{eff}}$ and $\log g$ values derived from the global analysis performed with \texttt{EXOFASTv2}. 
Those in red indicate the spectroscopic T$_{\mathrm{eff}}$ combined with the trigonometric $\log g$ obtained with \texttt{q$^{2}$}. Green symbols correspond to the location of both stars with the parameters of LE19. The solid lines mark Yonsei-Yale isochrones of the noted ages for [Fe/H] = 0.15 dex. 
 \label{fig.isochrones}}
\end{figure}

Additionally, we derived projected rotational velocities (v$\sin i$) based on the spectral synthesis of relatively isolated iron lines, as is detailed in \citet{Jofre2020}. In this case, however, we adopted the calibration of \citet{Valenti2005} to determine macroturbulence. We find that both objects have similar low velocities: v$\sin i$ = 3.96 $\pm$ 0.30 km s$^{-1}$ and v$\sin i$ = 3.61 $\pm$ 0.48 km s$^{-1}$, for WASP-160A and B, respectively.

\subsection{Consistency checks on T$_{\mathrm{eff}}$} \label{check.Teff}

To check the reliability of the derived spectroscopic T$_{\mathrm{eff}}$ values, we performed a series of independent tests. First, we computed photometric temperatures for both stars using the metallicity-dependent T$_{\mathrm{eff}}$-color calibrations formulae by \citet{Casagrande2010} using our highest precision [Fe/H] values from Table \ref{table.spectroscopic.parameters} and the ($B-V$), ($V-J$), ($V-H$), ($V-K_{S}$), and ($J-K_{S }$) colors calculated from the photometric data listed in Table \ref{table.all.stellar.parameters}. The average resulting values are T$_{\mathrm{eff}}$ = 5393 $\pm$ 45 K for WASP-160A and T$_{\mathrm{eff}}$ = 5179 $\pm$ 58 K for WASP-160B, which not only are in good agreement with the spectroscopic determinations but also keep a similar $\Delta$T$_{\mathrm{eff}}$ value.

As another test, we computed T$_{\mathrm{eff}}$ from EWs line strength ratios using the \texttt{T$_{\mathrm{eff}}$-LR} code\footnote{\url{http://www.astro.up.pt/~sousasag/ares/line_ratiopick2.php}} as described in \citet{Jofre2020}. In excellent agreement with the results from the differential spectroscopic analysis, we obtained  T$_{\mathrm{eff}}$ = 5400 $\pm$ 16 K for WASP-160A and T$_{\mathrm{eff}}$ = 5208 $\pm$ 18 K for WASP-160B. We also obtained T$_{\mathrm{eff}}$ from the analysis of the Spectral Energy Distribution (SED) based on the latest version of the \texttt{Virtual Observatory SED Analyzer}\footnote{\url{http://svo2.cab.inta-csic.es/theory/vosa/}} \citep[\texttt{VOSA},][]{Bayo2008} as detailed in \citet{Jofre2020}. The effective temperatures computed from VOSA are T$_{\mathrm{eff}}$ = 5400 $\pm$ 50 K and T$_{\mathrm{eff}}$ = 5200 $\pm$ 60 K for the components A and B, respectively, which once again are consistent with our spectroscopic estimates. 

\subsection{Stellar mass, radii, and ages} \label{sec.other.parameters}

Following \citet{Ramirez2014}, we derived stellar mass M$_{\mathrm{\star}}$, radius R$_{\mathrm{\star}}$, and ages $\tau_{\mathrm{\star}}$ using probability distribution functions with Yonsei-Yale (YY) stellar evolution models \citep{Demarque2004}. This was carried out via the \texttt{q$^{2}$} pipeline, which has been extensively tested in the literature \citep[e.g.,][]{Ramirez2014, Melendez2017, Bedell2017, Liu2020}.
As input, we used \textit{Gaia} DR2 parallaxes \citep{Gaia2018}\footnote{Parallaxes were corrected for the 82 $\mu$arcsec offset found by \citet{Stassun2018}.}, V magnitudes, and the spectroscopic T$_{\mathrm{eff}}$ and [Fe/H] values derived above (third block of Table \ref{table.spectroscopic.parameters}). For WASP-160A, this code returned M$_{\mathrm{\star}}$ = 0.946 $\pm$ 0.018 M$_{\mathrm{\odot}}$, R$_{\mathrm{\star}}$ = 0.934 $\pm$ 0.020 R$_{\mathrm{\odot}}$, and $\tau_{\star}$= 6.067 $\pm$ 1.689 Gyr, whilst for WASP-160B, we found M$_{\mathrm{\star}}$ = 0.881 $\pm$ 0.014 M$_{\mathrm{\odot}}$, R$_{\mathrm{\star}}$ = 0.860 $\pm$ 0.014 R$_{\mathrm{\odot}}$, and $\tau_{\star}$= 7.440 $\pm$ 1.597 Gyr. This code also provides the trigonometric gravities, which allow us to perform a consistency check on the spectroscopic $\log g$ values. \texttt{q$^{2}$} yields $\log g$ = 4.472 $\pm$ 0.026 dex  and $\log g$ = 4.513 $\pm$ 0.021 dex for WASP-160A and B, respectively. Both values are in agreement, within the errors, with the estimates found from the spectroscopic equilibrium.

We also derived M$_{\mathrm{\star}}$, R$_{\mathrm{\star}}$, and $\tau_{\mathrm{\star}}$ of both stars using the \texttt{EXOFASTv2} software package\footnote{https://github.com/jdeast/EXOFASTv2} \citep{Eastman2013, Eastman2019}. Here, the stellar models are simultaneously constrained by our newly derived effective temperature and metallicity from Table~\ref{table.spectroscopic.parameters}, the SED constructed from the catalog broadband photometry (see Table~\ref{table.all.stellar.parameters} and Figure \ref{fig.SED}), the \textit{Gaia} DR2 parallax \citep[also in Table~\ref{table.all.stellar.parameters};][]{Gaia2018}, and extinction in the direction of WASP-160 \citep[A$_V$ = 0.0308 $\pm$ 0.0015 mag;][]{Schlafly2011}. We set Gaussian priors for the parameters for which we have independent constraints, namely the T$_{\mathrm{eff}}$, [Fe/H], parallax and extinction, with widths corresponding to their one-sigma uncertainties. We set an upper limit to the extinction of 0.04 to avoid unrealistic solutions. Additionally, we used as starting points in $\log{g}$, M$_{\mathrm{\star}}$, R$_{\mathrm{\star}}$, and $\tau_{\mathrm{\star}}$, the values  derived with \texttt{q$^{2}$}, and since they are not independently constrained, we do not restrict them in the fitting. The stellar physical properties are then derived by simultaneously fitting a single star SED and either MIST \citep{Choi2016} or YY stellar models. For WASP-160B, an additional constraint is provided by the stellar density derived from the transit light curve data (see Section \ref{sec.planetary}). Thus, this code also allows us to make a consistency check both on the spectroscopic T$_{\mathrm{eff}}$ and $\log g$ derived above. Using the YY models, for WASP-160A \texttt{EXOFASTv2} returned T$_{\mathrm{eff}}$ = 5396 $\pm$ 19 K, $\log g$ = 4.449 $\pm$ 0.023 dex, M$_{\mathrm{\star}}$ = 0.926 $^{+0.024}_{-0.022}$ M$_{\mathrm{\odot}}$, R$_{\mathrm{\star}}$ = 0.950 $^{+0.018}_{-0.019}$ R$_{\mathrm{\odot}}$, and $\tau_{\star}$= 8.0 $^{+2.5}_{-2.6}$ Gyr, whilst for WASP-160B, we found T$_{\mathrm{eff}}$= 5202 $\pm$ 15 K, $\log g$ = 4.501 $\pm$ 0.016 dex, M$_{\mathrm{\star}}$ = 0.879 $^{+0.022}_{-0.020}$ M$_{\mathrm{\odot}}$, R$_{\mathrm{\star}}$ = 0.872 $^{+0.013}_{-0.012}$ R$_{\mathrm{\odot}}$, and $\tau_{\star}$= 8.2 $^{+2.8}_{-3.0}$ Gyr. These values are almost identical to those obtained using the MIST grids and, moreover, they are in excellent agreement with those from \texttt{q$^{2}$}.

As final values of mass, radii, and ages we adopted
the results provided by \texttt{EXOFASTv2}\footnote{These parameters correspond to the median in the MCMC final distributions and the one-sigma uncertainties to the 68\% confidence intervals.} which are listed in Table~\ref{table.all.stellar.parameters}. Here, we also have included the adopted differential atmospheric parameters, which for WASP-160A are those using the Sun as reference whilst the parameters for WASP-160B are relative to WASP-160A (third block of Table \ref{table.spectroscopic.parameters}).   

The small difference between the spectroscopic $\log g$  and T$_{\mathrm{eff}}$ values and those derived using photometry and parallax information, all within 1$\sigma$, can be seen in Figure \ref{fig.isochrones}. Here we plot the locations of both stars on the T$_{\mathrm{eff}}$ versus $\log g$ diagram, along with YY theoretical isochrones. Assuming that the stars are bound and coeval, their ages should be consistent and fall approximately on the same isochrone. Our spectroscopic $\log g$ values would make both stars slightly older than $\sim$ 9 Gyr, whilst if we consider the position given by the trigonometric $\log g$ values and our spectroscopic T$_{\mathrm{eff}}$ obtained with \texttt{q$^{2}$}, both stars are located over the 6--7 Gyr isochrones. For the results derived from the \texttt{EXOFASTv2}, the stars fall on the same $\sim$ 8 Gyr isochrone. However, given the obtained uncertainties, all ages would be consistent with those of thin disk stars \citep[$\lesssim$9 Gyr, e.g.,][]{Bernkopf2001, Nissen2004, Kilic2017}, which is in agreement with the Galaxy population membership derived for the stars in Section \ref{sec.abundances}. As we will see later in Section \ref{impact.parameters}, the small difference in the $\log g$ values does not affect our relative abundance ratios in any significant way. In agreement with the location of both stars on the T$_{\mathrm{eff}}$ - $\log g$ plane, based on the calibration of \citet{Pecaut2013} and our derived stellar parameters, we estimated that WASP-160A and WASP-160B would correspond to G8 and K0.5 dwarf stars. 

The only previous estimations of stellar mass, radius, and age for the WASP-160 stars were reported by LE19. They used their spectroscopic fundamental parameters, the stellar density derived from the transit light curve (only for WASP-160B), \textit{Gaia} photometry, and the distance as initial constraints to obtain stellar masses, radii, and ages through PARSEC stellar evolution models \citep{Bressan2012}. All the parameters agree with our results within the errors. 


\subsection{Elemental abundances} \label{sec.abundances}
Adopting the spectroscopic atmospheric parameters presented in the third block of Table \ref{table.spectroscopic.parameters} as the first choice, we measured the abundances of 25 elements other than iron for the WASP-160 stars. The abundances were derived from EWs using the curve-of-growth approach with the MOOG code (abfind driver) via the \texttt{q$^{2}$} code. The EWs were measured following the same procedure employed for the iron lines (see Section \ref{sec.atmospheric.parameters}). The line list and atomic parameters adopted were taken from \citet{Ramirez2014} and \citet{Melendez2014}. We also included near-IR lines for C from \citet{Amarsi2019} and for N from  \citet{Takeda2001} and \citet{Ecuvillon2004}. The abundances of Sc, Ti, and Cr (in addition to Fe) were obtained from both neutral and singly-ionized species, whilst for Y, Zr, Ba, and Eu only singly-ionized species are available. For the rest of the elements, only neutral species were used. Hyperfine splitting was taken into account for V, Mn, Co, Cu, Rb, Y, Ba and Eu. The O abundance was computed from the 7771-5 {\AA} IR triplet, adopting the non-LTE corrections by \citet{Ramirez2007}. The line list and atomic parameters used in this work  along with the EWs measured in the WASP-160 stars and our solar spectrum are listed in Table \ref{table.lines} of the Appendix \ref{appendixA}. The line-by-line differential abundances relative to solar ([X/H]) and those using WASP-160A as the reference star ($\Delta$[X/H]$_{B-A}$), along with the errors, are listed in Table \ref{table.abundances} (Appendix \ref{appendixA}). These errors are computed adding in quadrature both the line-to-line scatter\footnote{For species with only one line (K \textsc{i}, Zr \textsc{ii}, and Eu \textsc{ii}), we conservatively adopted the largest line-to-line scatter obtained for species with more than three lines available.} and the errors propagated from each parameter uncertainty as in \citet{Ramirez2015}. Our errors in relative abundances, $\Delta$[X/H]$_{B-A}$, vary from one species to another in the 0.005--0.030 dex range. In particular, the median error is 0.017 dex and 0.008 dex for elements with T$_{c}$ $\lesssim$ 900 K and for those with T$_{c}$ $>$ 900 K, respectively. 

In both stars the lithium line (Li \textsc{i}) at 6707.8 {\AA} is not detected (Figure \ref{fig.spectra}, right) and therefore we can only derive an upper limit of A(Li) $\lesssim$ 0.17 dex and A(Li) $\lesssim$ 0.6 dex for WASP-160A and B, respectively, by performing a synthesis analysis. This result is in agreement with the non-detection of lithium reported by LE19 for WASP-160B, and supports the relatively old ages derived in Section \ref{sec.other.parameters}. The low abundance of lithium also might be caused by the relatively low temperature of WASP-160A and B since stars with T$_{\mathrm{eff}}$ below 5500 K have A(Li) $\lesssim$ 0.5 dex \citep{Ramirez2011, Aguilera2018}, as a consequence of having deeper convective zones.

Considering the relatively old age of the WASP-160 stars, we determined their Galactic population membership. First, following the procedure detailed in \citet{Jofre2015}, we derived the Galactic space-velocity components (U, V, W) listed on Table \ref{table.all.stellar.parameters} using the radial velocities measured from the GRACES spectra and the proper motions and parallaxes from \textit{Gaia} DR2\footnote{We corrected the \textit{Gaia} parallaxes for the 82 $\mu$arcsec suggested by \citet{Stassun2018}.}. From these space-velocity components and the membership formulation by \citet{Reddy2006}, we found that the probability of belonging to the thin disk population is $\sim$99\% for both stars. The $\alpha$-elemental abundances of WASP-160A ([$\alpha$/Fe]= 0.02 dex) and WASP-160B ([$\alpha$/Fe]= 0.04 dex) appear to be consistent with the kinematic thin-disk membership of the system.

Strictly speaking the components of the WASP-160 stellar pair, with consistent astrometry and near-identical systemic RVs, are by definition comoving stars and, in principle, there is no guarantee that they are gravitationally bound \citep[see e.g.,][]{Oh2017, Simpson2019, Ramirez2019}. However, their chemical compositions along with their same old ages ($\sim$8 Gyr), which are rare for thin disk objects, suggest that a lucky alignment of these two comoving stars or a capture scenario is unlikely. Moreover, following \citet{Andrews2017} and \citet{Ramirez2019}, we examined the location of the WASP-160 pair in the plot of projected separation versus total velocity difference \citep[see Fig. 7 in][]{Ramirez2019}. With a projected separation of  $s$ = 8060 AU between the components ($\log s$ = 3.90 AU) and a difference in total velocity of $\Delta V$ = 0.02 km $s^{-1}$ ($\log \Delta V$ $\sim$ $-$1.70 km $s^{-1}$)\footnote{$\log \Delta V$ $\sim$ $-$1.3 km $s^{-1}$, if we consider the velocity values of Gaia DR2.}, WASP-160 fall safely within the limits where a system can be considered as physically bound and hence a \textit{true} binary.

\section{Chemical differences between WASP-160A and B}\label{sec.differences.abundances}
\begin{figure*}[ht!]
\centering
\includegraphics[width=.75\textwidth]{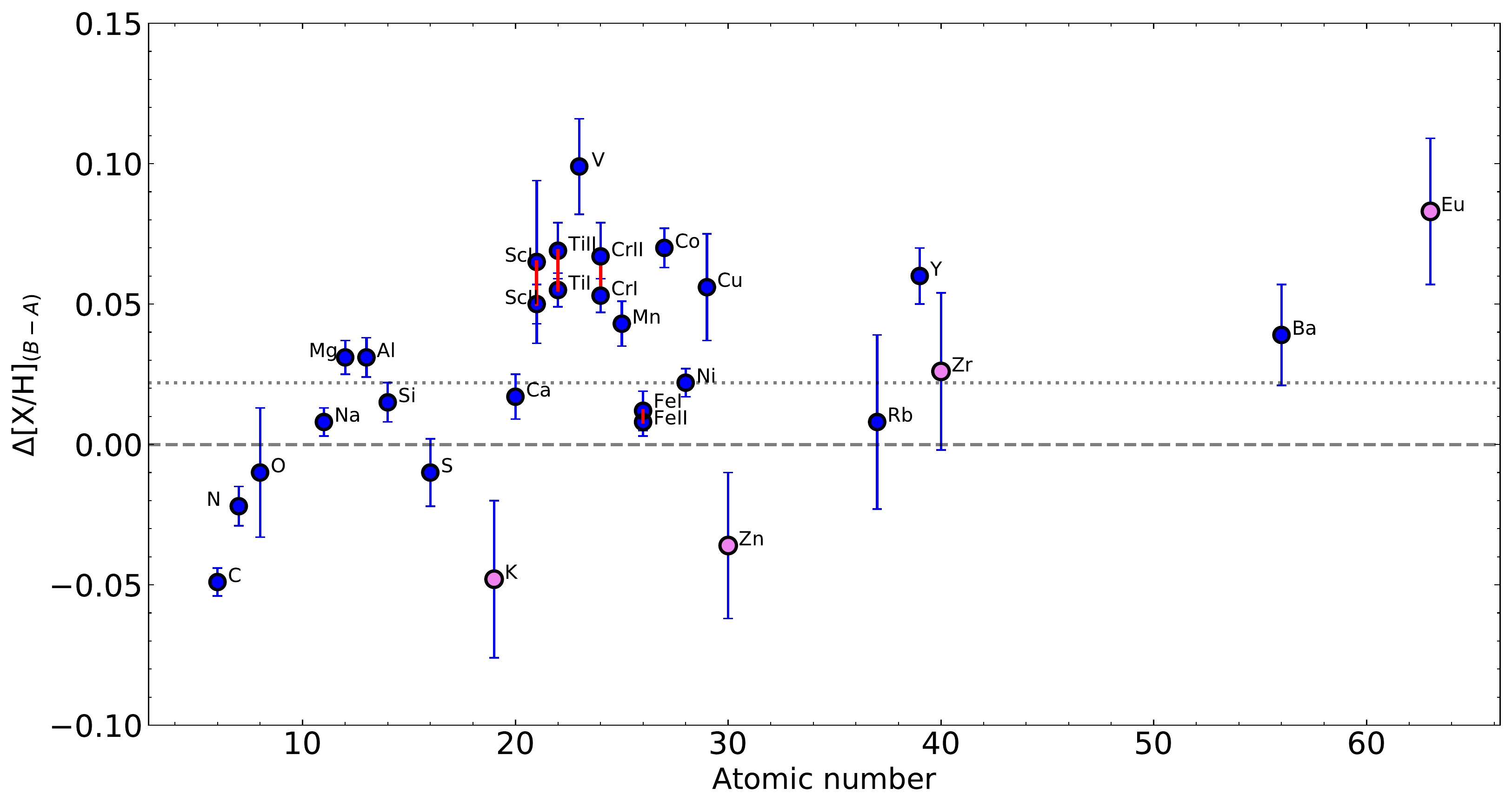}
\caption{Elemental abundance differences between WASP-160B and WASP-160A as a function of atomic number. The dashed line corresponds to identical composition and the dotted line represents the weighted average of $\Delta$[X/H]$_{B-A}$ considering all the elements but those measured from only one line (see text for more details). Red vertical lines connect two species of the same chemical element (e.g., Sc \textsc{i} and Sc \textsc{ii}) and  pink circles show the species measured from one line only (K \textsc{i}, Zn \textsc{i}, Eu \textsc{ii}, and Zr \textsc{ii}). \label{fig.abun.vs.Z}}
\end{figure*}

Figure \ref{fig.abun.vs.Z} shows the derived differential abundances of WASP-160B relative to WASP-160A, $\Delta$[X/H]$_{B-A}$, as a function of atomic number. We can see that WASP-160B is more metal-rich in almost all elements relative to its binary companion WASP-160A. Conservatively, if we exclude the abundances of those elements measured from only one line (i.e., K, Zn, Eu, and Zr), we find a weighted average (and weighted standard deviation) of $\Delta$[X/H]$_{B-A}$ = 0.022 $\pm$ 0.006 dex (i.e., an excess detected at the $\sim$4$\sigma$ level). Hereafter, we discuss our results ignoring the K, Zn, Eu, and Zr abundances\footnote{The results, however, remain almost unchanged if we include these elements measured from one line only.}. 

\begin{figure*}[ht!]
\centering
\includegraphics[width=.75\textwidth]{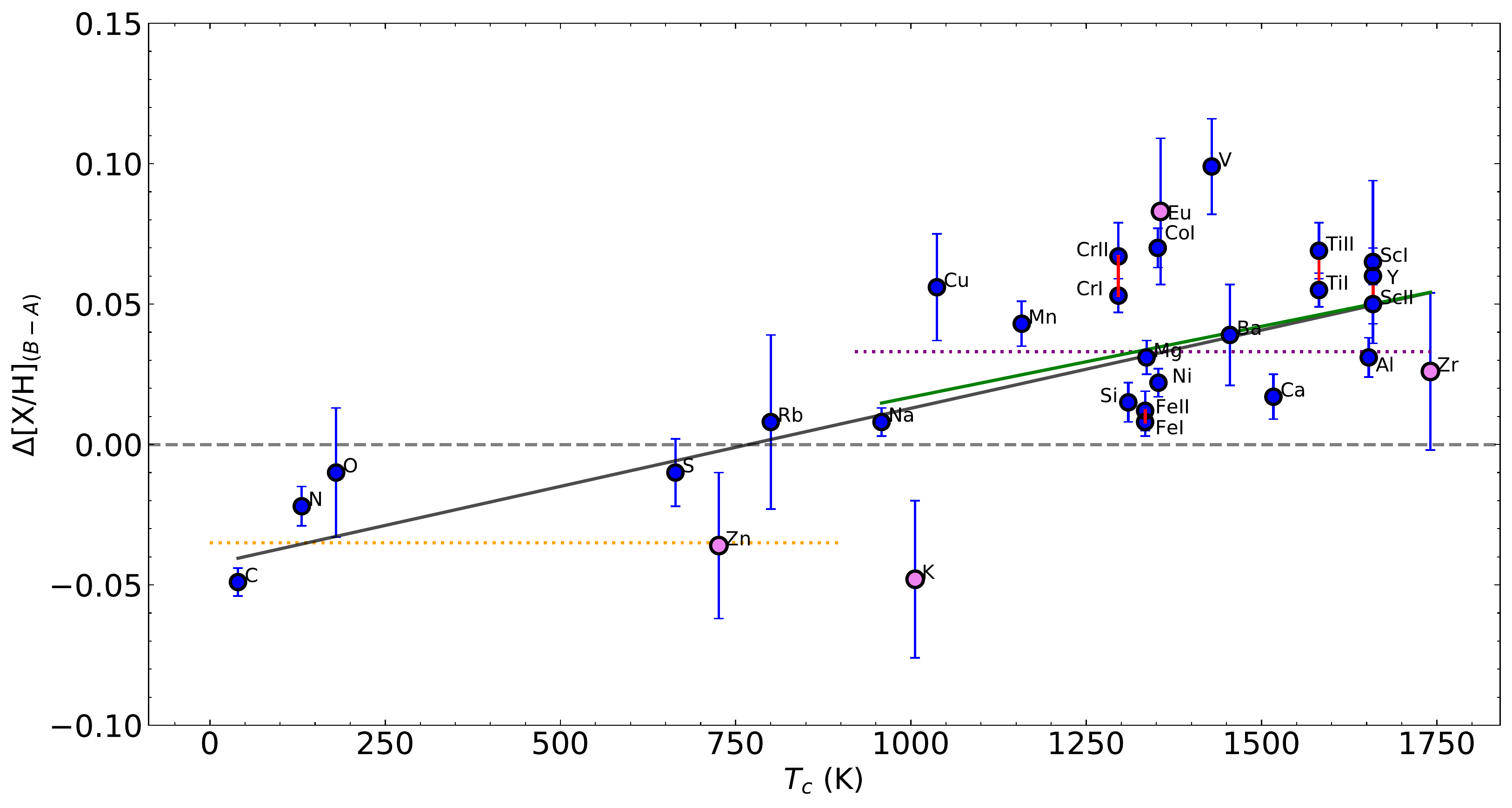}
\caption{Chemical composition difference between WASP-160B and WASP-160A versus condensation temperature. The solid black and green lines are the weighted linear least-squares fit to all the elements and refractory elements measured from more than one line, respectively. The horizontal dotted orange and violet lines represent the weighted average for the volatile and refractory elements, respectively. As before, the dashed line corresponds to identical composition, and the solid red vertical lines connect two species of the same chemical element and  pink circles show the species measured from one line only. \label{fig.Tcond.todo}}
\end{figure*}

In Figure \ref{fig.Tcond.todo}, we show the derived differential elemental abundances of WASP-160B relative to WASP-160A versus the 50\% condensation temperature (T$_{c}$) from \citet{Lodders2003} for solar-composition gas. The $\Delta$[X/H]$_{B-A}$ seems to correlate with T$_{c}$, with the lowest values for the volatile elements like carbon around $\sim$$-$0.05 dex, and the highest up to $\sim$0.1 dex for vanadium, the refractory element with the highest T$_{c}$ in our list. In particular, we note that the Saturn-mass planet-host star WASP-160B seems to be depleted in volatiles (elements with T$_{c}$ $\lesssim$ 900 K), with a weighted average and standard deviation of $\Delta$[X/H]$_{B-A}$ = $-$0.035 $\pm$ 0.008 dex (i.e., a deficiency detected at the 5$\sigma$ level) and enhanced in refractories (elements with T$_{c}$ $>$ 900 K) with a weighted average and standard deviation of $\Delta$[X/H]$_{B-A}$ = +0.033 $\pm$ 0.005 dex (i.e., an excess detected at the 7$\sigma$ level) relative to its companion WASP-160A. To date, no planet has been detected around WASP-160A. 

We tested three models on the $\Delta$[X/H]$_{B-A}$ versus T$_{c}$ data to investigate a possible correlation. First, a zero-slope ($m$=0) and zero-intercept fit ($b$=0; the black-dashed line in Figure \ref{fig.Tcond.todo}) results in a $\chi_{r} ^{2}$ value\footnote{$\chi_{r} ^{2} = \chi^{2}/(n-d)$, where $n$ is the number of data-points,  $d$ is the number of free parameters in the model, $\chi^{2} = \Sigma [\Delta[X/H]-\Delta[X/H]_{mod}]^2/$Error$^2_{obs}$, and $\Delta[X/H]_{mod}$ represents the differential abundance predicted by the model.} of 27.9, a mean scatter of $\sigma_{f}$= 0.033, and a Bayesian Information  Criterion\footnote{BIC$ = \chi^{2} + k_{f} \log N_{p}$ ; where $k_{f}$ is the number
of free parameters in the model and $N_{p}$ is the number of data-points  \citep{Schwarz1978}.} value of BIC = 699. Second, a weighted linear fit to the data with $m$=0 and free intercept gives $b$= 0.022 $\pm$ 0.006 dex with $\chi_{r} ^{2}$ = 19.9, $\sigma_{f}$= 0.033, and BIC= 481. Finally, a weighted\footnote{By 1/$\sigma^{2}_{\Delta[X/H]_{B-A}}$} linear fit to the data with $m$ and $b$ as free parameters results in an intercept of $b$=$-$0.042 $\pm$ 0.004 dex and a positive slope of (5.56 $\pm$ 0.79) $\times 10^{-5}$  dex K$^{-1}$, with $\chi_{r} ^{2}$ = 6.6, $\sigma_{f}$= 0.021, and BIC = 158. Considering the $\chi^{2}$, $\sigma_{f}$, and BIC values obtained for each model, we find that the linear fit with unconstrained slope and intercept provides the best fit to the data in comparison with the other two models\footnote{A comparison between the two best models (the one with $m$ and $b$ as free parameters and that with free $b$ but fixed $m$) gives $\Delta$BIC = 324. According to \citet{Kass1995}, such a difference strongly supports the positive slope model.}. 

Considering all elements (both volatiles and refractories), the analysis above supports a non-negligible slope for the $\Delta$[X/H]$_{B-A}$ versus T$_{c}$ data, with a significance of the slope at the $\sim$7$\sigma$ level. The Pearson's \textit{r} correlation test gives in this case a coefficient of \textit{r} $\sim$ 0.76. Moreover, following \citet{Maldonado2016}, we carried out 100,000 bootstrap Monte Carlo (MC) simulations to estimate the probability that uncorrelated random data sets could reproduce the observed trend in Figure \ref{fig.Tcond.todo}, and hence the value of the slope. We found that this probability is only $\sim$4.6\%, which gives additional support to a significant correlation between the differential abundances and T$_{c}$.  If we only consider the behavior of the refractory elements, the data still favor a model with a non-zero positive slope over the other models. However, in this case, the slope of the weighted linear fit to refractory abundances versus T$_{c}$ has a modest statistical significance ($\sim$2$\sigma$). 

\subsection{Impact of the Adopted Stellar Parameters}\label{impact.parameters}

\citet{Teske2015} found that chemical differences between the stellar components of a binary system, and therefore the T$_{c}$ trends, are sensitive to the chosen stellar parameters. Following the exercise presented in \citet[][their Section 4.7]{Ramirez2015}, here we show how the $\Delta$[X/H]$_{B-A}$ versus T$_{c}$ trend observed in Figure \ref{fig.Tcond.todo} depends on the adopted fundamental parameters.

\begin{figure}[ht!]
\includegraphics[width=.45\textwidth]{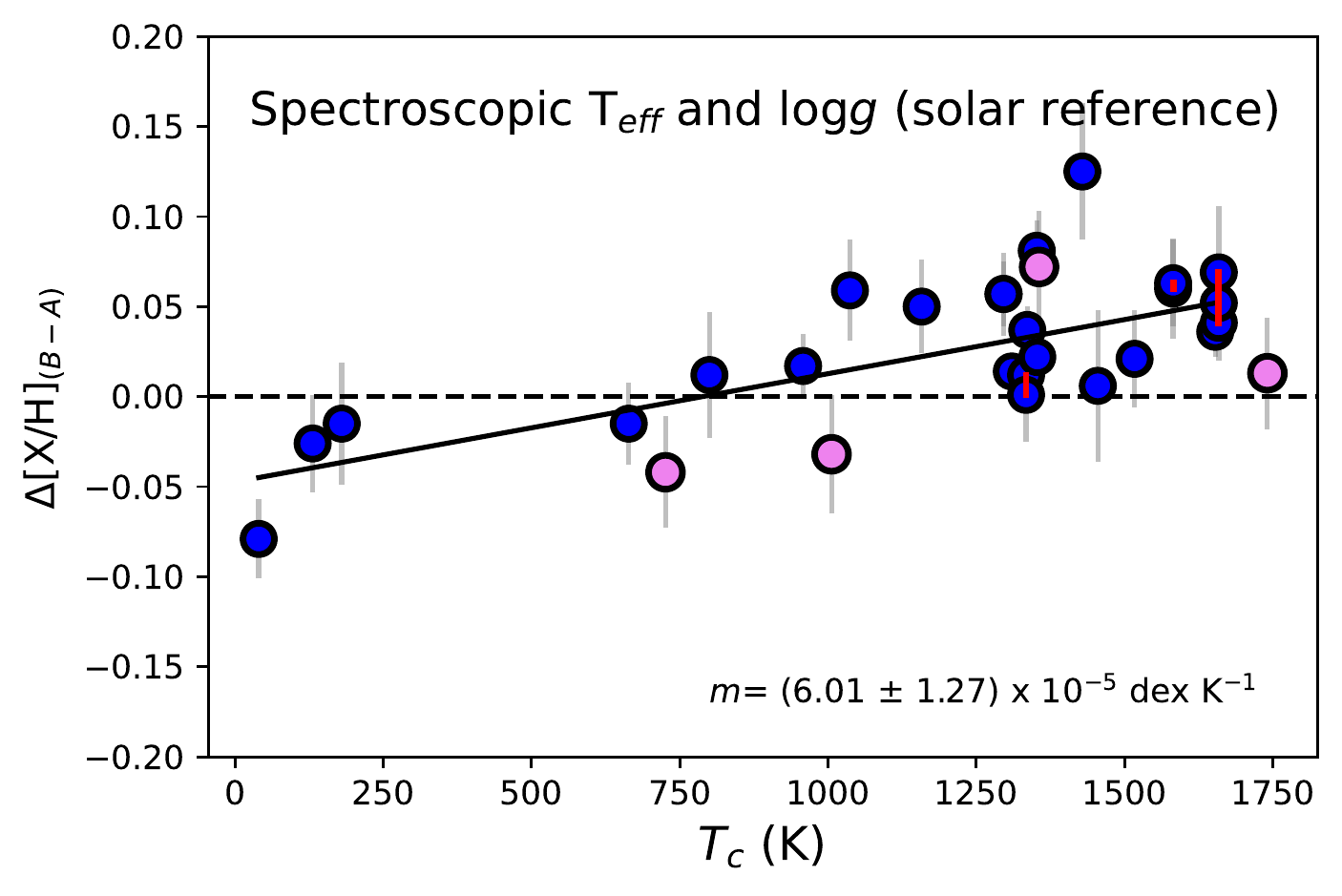}
\includegraphics[width=.45\textwidth]{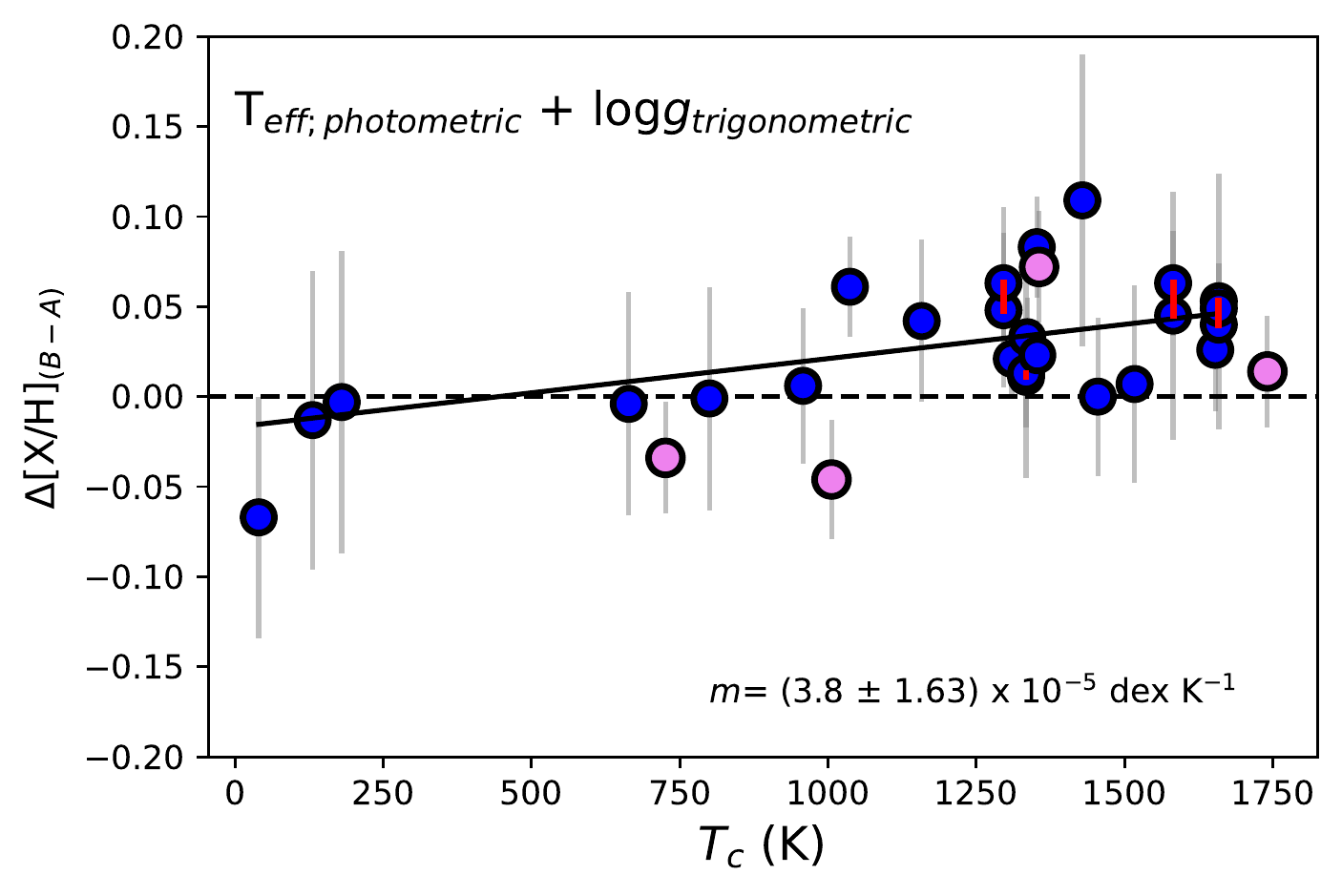}
\includegraphics[width=.45\textwidth]{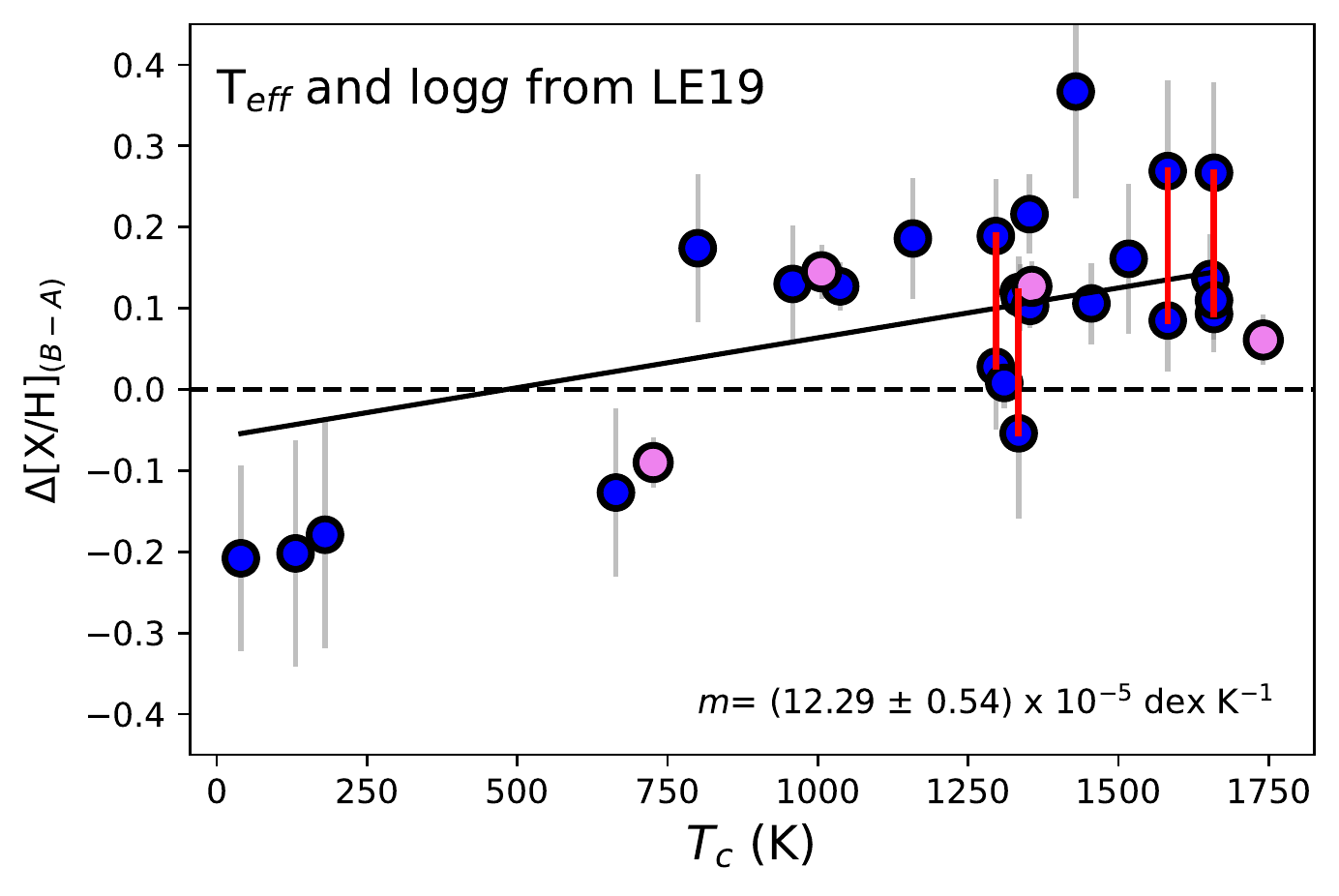}
\caption{Same as in Figure \ref{fig.Tcond.todo}, but using different choices of stellar parameters for WASP-160A and WASP-160B. The derived slope values of the fits are presented in the bottom-right corner of each panel.} \label{fig.Tcond.set.parameters}
\end{figure}

Using our EWs measurements, we started by recomputing the $\Delta$[X/H]$_{B-A}$ values using different sets of stellar parameters. First, we recalculated the abundances from our spectroscopic parameters of the WASP-160 stars computed using the Sun as reference for both stars (second block of Table \ref{table.spectroscopic.parameters}). The abundances versus T$_{c}$ correlation with this set of parameters, shown in the first panel of Figure \ref{fig.Tcond.set.parameters}, is very similar to the one obtained with our preferred set of parameters. Here, the offset of volatiles is $-0.033$ dex (weighted average) whilst the one for refractories is +0.035 dex. These values are very similar to those obtained using WASP-160A as reference. In this case, it is interesting to note that although the $\Delta$T$_{\mathrm{eff}}$ is basically the same (only 3 K smaller), a smaller $\Delta \log g$ of only 0.02 dex in comparison with a difference of 0.05 dex for the original parameters, does not produce a significant impact on the observed correlation.         

The middle panel in Figure \ref{fig.Tcond.set.parameters} shows the chemical pattern that results when we use the temperatures derived from photometric calibrations and the trigonometric $\log g$ values computed with \texttt{q$^{2}$}. Again, since the B$-$A relative parameters are the same in comparison with the previous case, a very similar and significant $\Delta$[X/H]$_{B-A}$ versus T$_{c}$ correlation is obtained although with a slightly smaller slope (3.8 $\pm$ 1.6 $\times 10^{-5}$  dex K$^{-1}$). Analogous results are obtained using the T$_{\mathrm{eff}}$ and $\log g$ values derived from \texttt{EXOFASTv2} (not plotted).

The bottom panel shows the abundances derived using the parameters from LE19, which are listed in the first block of Table \ref{table.spectroscopic.parameters}. In this case, as we mentioned in Section \ref{sec.atmospheric.parameters}, there is a relatively discrepancy between our stellar parameters and those derived by LE19, but more importantly a larger difference in the B$-$A relative parameters. In particular, LE19 suggest $\Delta$T$_{\mathrm{eff}}$ = 0 K,  $\Delta \log g$ = 0.05 dex, and $\Delta$[Fe/H] = 0.08 dex whilst for our preferred set we find $\Delta$T$_{\mathrm{eff}}$= 209 K,  $\Delta \log g$=0.05 dex, and $\Delta$[Fe/H] = 0.012 dex. Using the parameters of LE19, the offset between volatiles and refractories seems larger and, hence, causes a steeper correlation. However, the scatter is larger than the observed with the other choice of parameters. Moreover, with these  parameters, the abundances of elements obtained from both neutral and singly-ionized species (e.g., Fe, Ti, Cr, Sc, indicated with red lines) show the largest differences in comparison with those obtained with other sets of parameters. For example, the difference $\Delta$[Ti \textsc{i}/H]$_{B-A}$ $-$ $\Delta$[Ti \textsc{ii}/H]$_{B-A}$ is 0.184 dex in this case, but we find a difference of $-$0.014 dex with our parameters. Based on these results, we consider that our derived stellar parameters are more accurate than those presented by LE19 \citep[see also,][]{Ramirez2015}. 

\subsection{Effect of $\Delta$T$_{\mathrm{eff}}$} \label{effect.Teff}
In the same manner as \citet[][see their Figure 8]{Ramirez2015} and \citet{Saffe2017}, here we explore how different our T$_{\mathrm{eff}}$ values ($\Delta$T$_{\mathrm{eff}}$ in particular) should be to completely blur the T$_{c}$ trend and/or flip the sign of the slope.

\begin{figure*}[ht!]
\centering
\includegraphics[width=.45\textwidth]{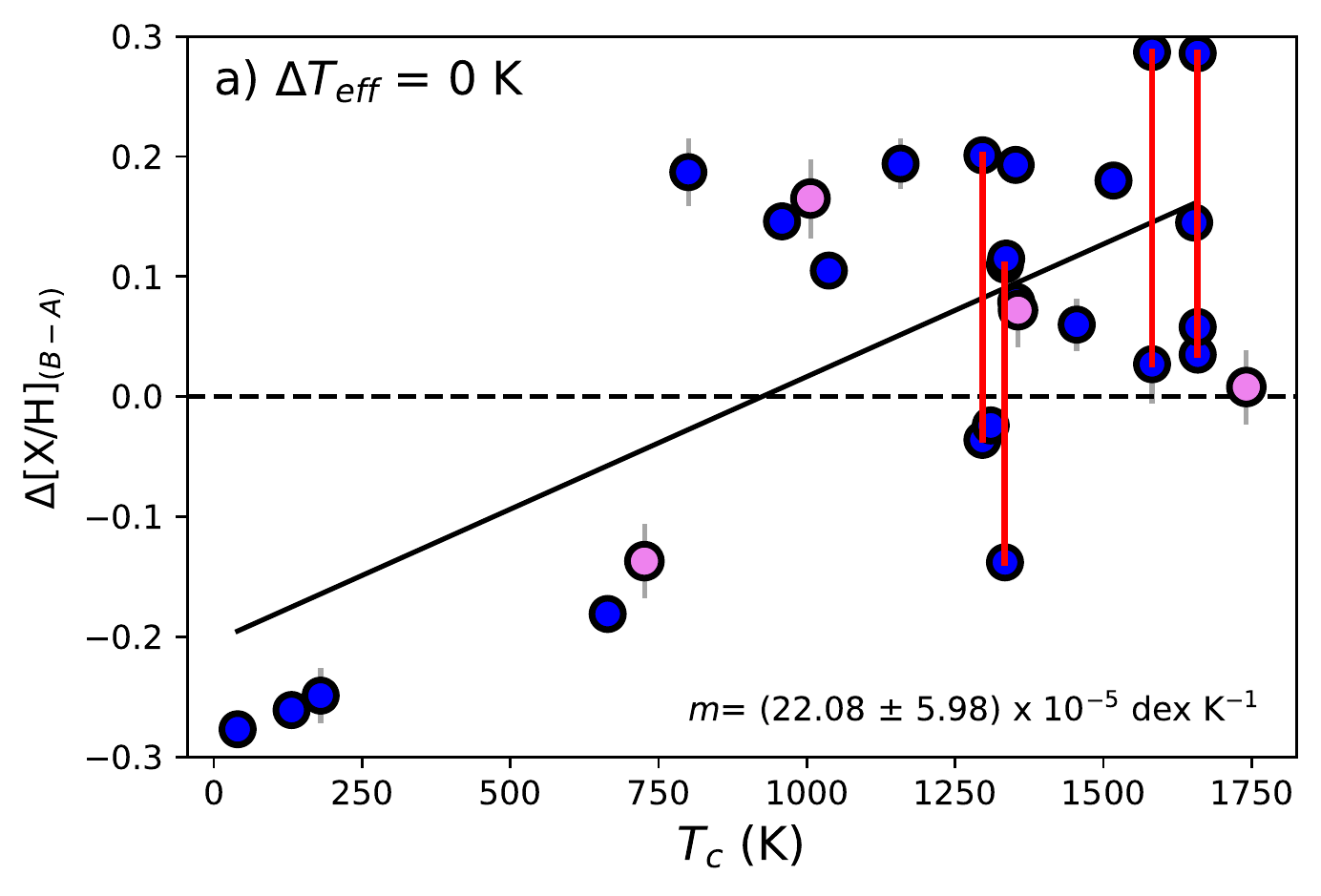}
\includegraphics[width=.45\textwidth]{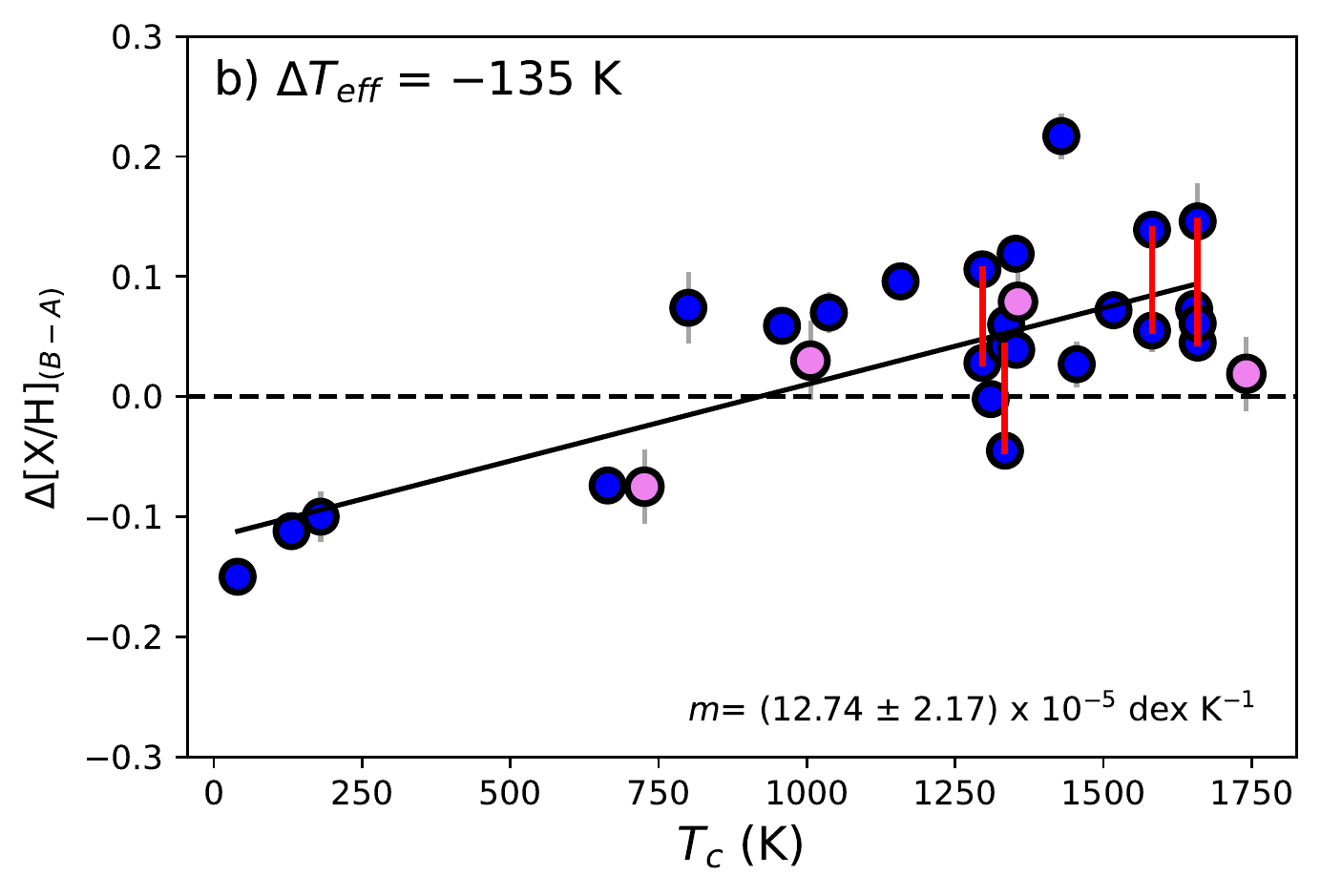}
\includegraphics[width=.45\textwidth]{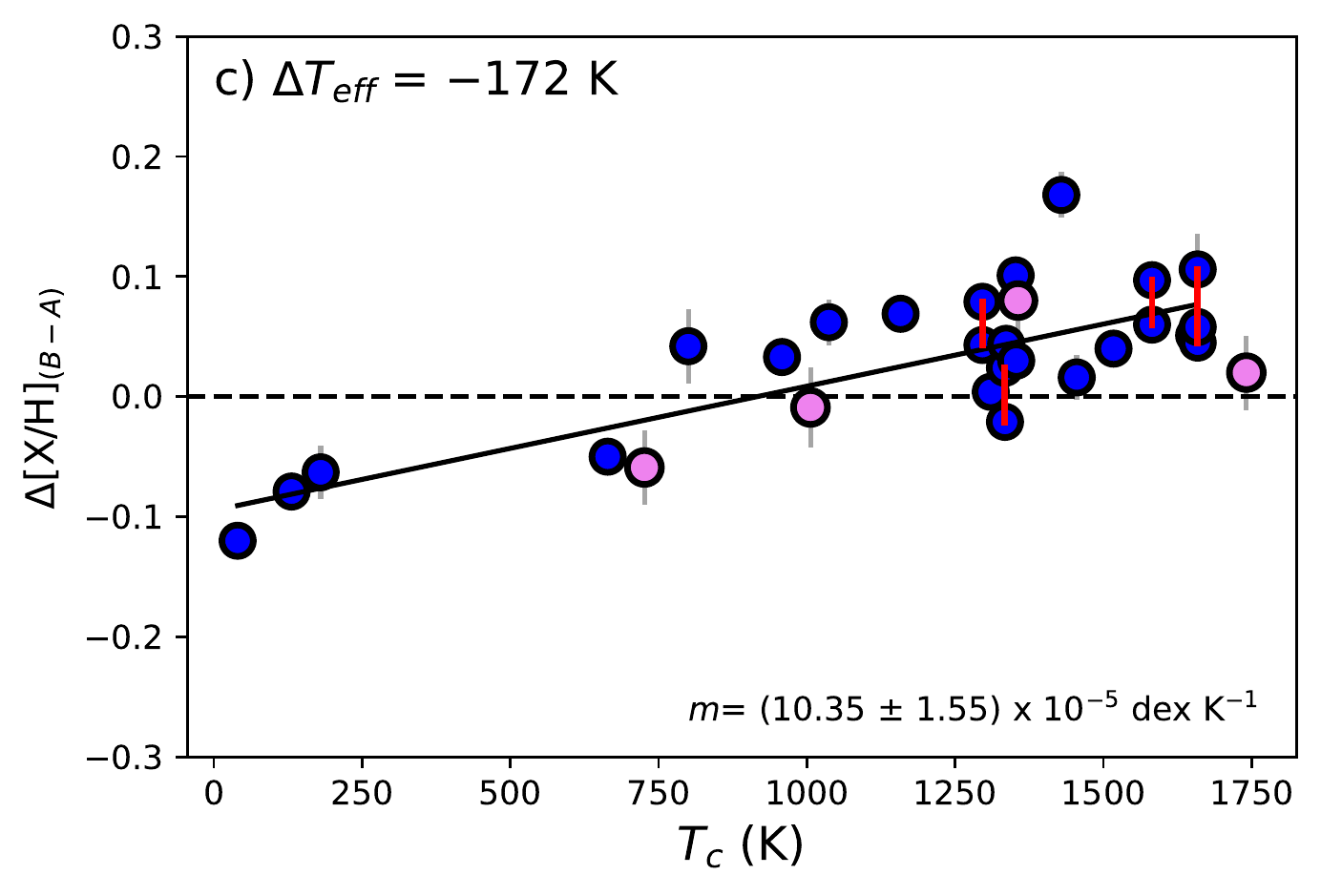}
\includegraphics[width=.45\textwidth]{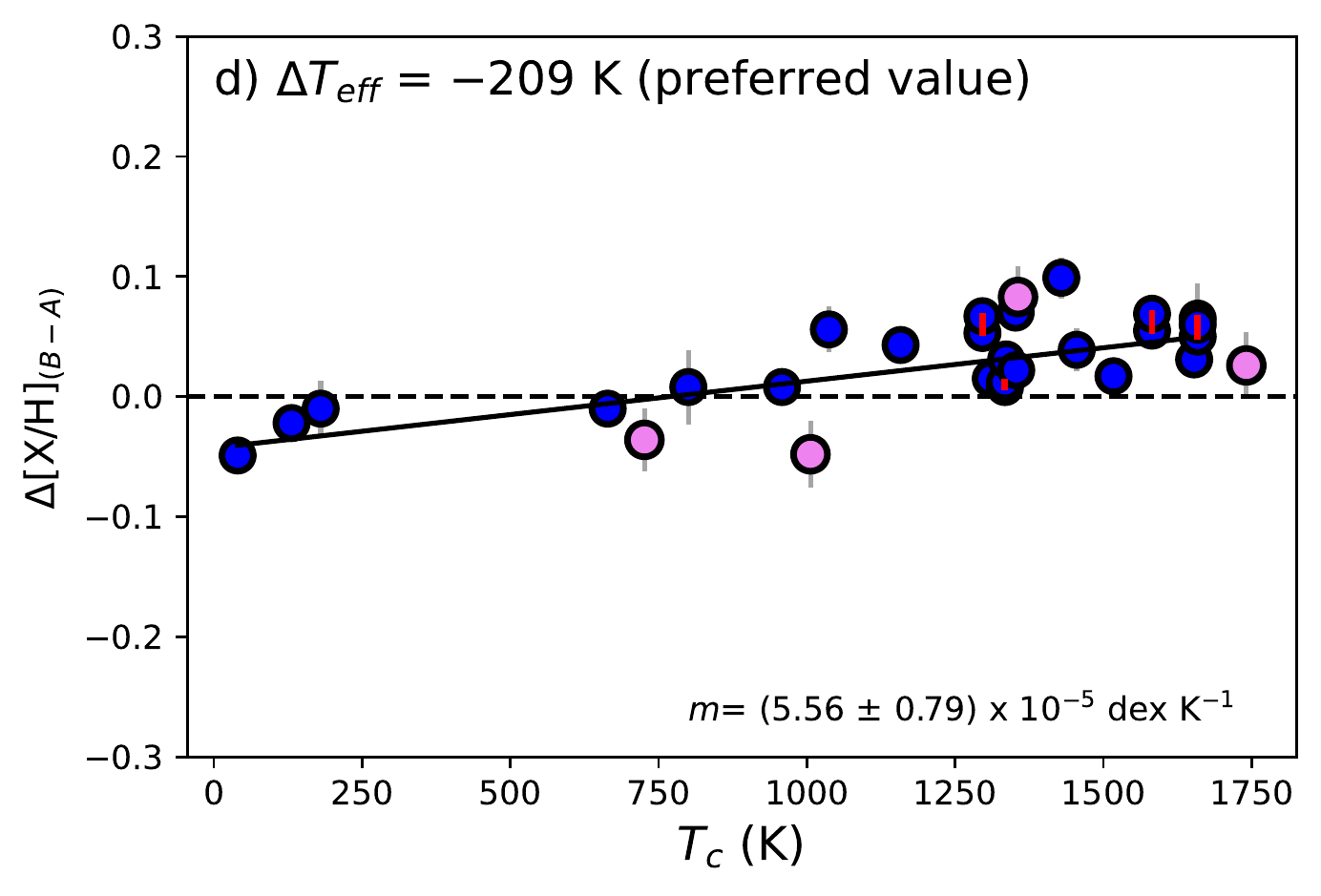}
\includegraphics[width=.45\textwidth]{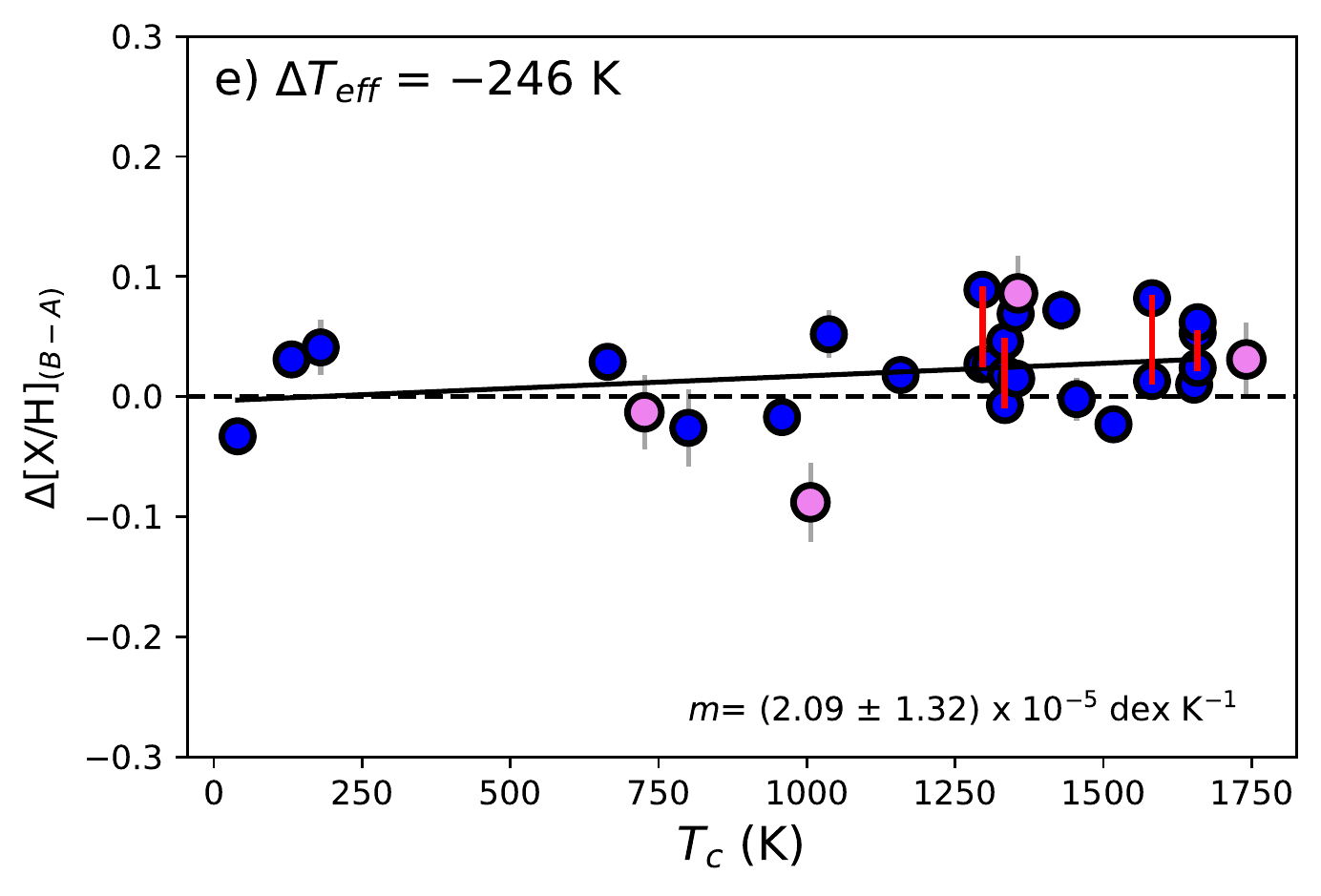}
\includegraphics[width=.45\textwidth]{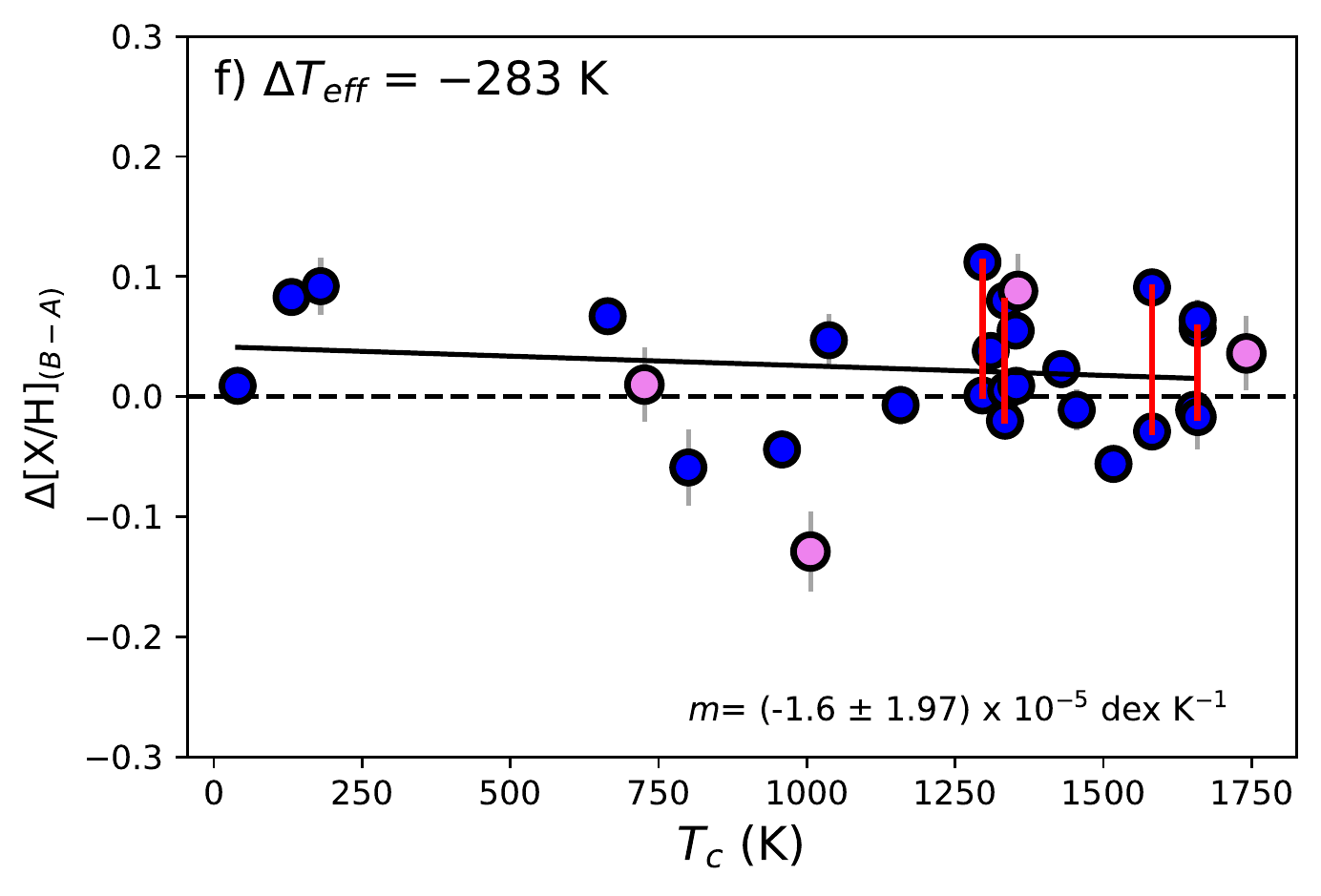}
\caption{Same as in Figure \ref{fig.Tcond.todo}, but considering different values of the B--A temperature difference, $\Delta$T$_{\mathrm{eff}}$ and keeping all other parameters constant. In this case, the panel with $\Delta$T$_{\mathrm{eff}}$ = $-$209 K is the same as Figure \ref{fig.Tcond.todo}. \label{fig.Tcond.delta.teff} The derived slope values of the fits are presented in the bottom-right corner of each panel.}
\end{figure*}

In Figure \ref{fig.Tcond.delta.teff} we show the abundances $\Delta$[X/H]$_{B-A}$ that result considering different values of $\Delta$T$_{\mathrm{eff}}$ (from $-$283 K to $-$135 K) in the sense B$-$A. Here, only the effective temperature difference is changed by altering the temperature of the WASP-160B, while all the other parameters are kept constant.

Starting from our preferred values in panel (d), we note that the $\Delta$[X/H]$_{B-A}$ versus T$_{c}$ correlation becomes steeper, and also noisier, as the $\Delta$T$_{\mathrm{eff}}$ value decreases (see panel c, b, and a). Moreover, the abundance differences for elements with neutral and singly-ionized species available (Fe, Cr, Ti, Sc) increases (see the filled circles connected by red vertical lines). This behavior reaches a maximum considering, as an additional case, that the two stars are twins in terms of T$_{\mathrm{eff}}$ (i.e., $\Delta$T$_{\mathrm{eff}}$ = 0 K), which is very similar to the observed result when using the parameters of LE19. Thus, the large difference between the T$_{\mathrm{eff}}$ of the components of the WASP-160 system could not lead by itself to the observed T$_{c}$ trend since the change in the slope goes in the opposite direction, i.e, as the stars become more similar in T$_{\mathrm{eff}}$ the T$_{c}$ correlation is steeper.

Analyzing the behavior of the T$_{c}$ trend at larger values of $\Delta$T$_{\mathrm{eff}}$ (i.e., less similar stars) beyond the preferred value can help us to explore when the correlation becomes marginal and/or changes its sign. Panels (e) and (f) in Figure \ref{fig.Tcond.delta.teff} show the results when we derive the abundances considering $\Delta$T$_{\mathrm{eff}}$ values of $-$246 K and $-$283 K. Again, it can be noticed that as we move away from our preferred value, the difference between the abundances of elements derived from both neutral and singly-ionized species increases. 

In particular, for the $\Delta$T$_{\mathrm{eff}}$= $-$283 K we find that the trend shows a slightly negative slope whilst for $\Delta$T$_{\mathrm{eff}}$= $-$246 K the trend exhibits a less steep slope (2.09 $\pm$ 1.32 $\times 10^{-5}$  dex K$^{-1}$ ) and, moreover, a linear fit with free values of $m$ and $b$ is no longer preferred against one with zero slope and free intercept. Nevertheless, in these two cases, WASP-160B remain chemically enhanced compared with WASP-160A. Moreover, for $\Delta$T$_{\mathrm{eff}}$= $-$246 K, WASP-160B is still enhanced (by $\sim$ 0.007 dex at the 3$\sigma$ level) relative to WASP-160A and the trend is still present but modest statistical significant ($\sim$2$\sigma$) when considering only refractories. 

To reach $\Delta$T$_{\mathrm{eff}}$ of  $\sim$$-$246 K, it would be required that simultaneously the T$_{\mathrm{eff}}$ of WASP-160A be overestimated and that of WASP-160B be underestimated almost by 2$\sigma$. However, in addition to the behavior of the abundances of elements derived from both neutral and singly ionized lines, the excellent agreement between the spectroscopic $\Delta$T$_{\mathrm{eff}}$ value and those derived from independent techniques, gives us no reason to assume that our $\Delta$T$_{\mathrm{eff}}$ is different from the preferred and adopted one. 

For these two last cases, and especially for that of $\Delta$T$_{\mathrm{eff}}$ = $-$246 K,  one might still wonder if exists a small $\Delta \log g$ value (within the total errors in $\log g$ for example) beyond our preferred value ($\Delta \log g$ = 0.05 dex) for which the discrepancy between the abundances of elements derived from both neutral and singly-ionized species decreases and reaches a minimum and what would happen with the trend in that case. To answer these questions, we repeated the experiment but now considering different values of $\Delta \log g$ in steps of 0.05 dex and keeping constant the other parameters. For $\Delta$T$_{\mathrm{eff}}$ = $-$246 K, we found that the abundance differences for elements with two species available start to decrease and reach a minimum when $\Delta \log g$ = $-$0.1 dex. As before, to reach this value of $\Delta \log g$ it would be required that simultaneously the $\log g$ of WASP-160A be underestimated and that of WASP-160B be overestimated in $\sim$2$\sigma$. Again, given the good agreement between the spectroscopic surface gravity values and those derived with other methods, we find no reason to assume that the $\Delta \log g$ can go beyond $\sim$$\pm$0.1 dex. Moreover, for the values of $\Delta \log g$ between 0 and $-$0.1 dex there is no significant change in the trend.  A very similar situation occurs for $\Delta$T$_{\mathrm{eff}}$ = $-$283 K, but in this case the minimum  abundance differences for elements with two species is reached for $\Delta \log g$ = $-$0.15 dex. 

Finally, we acknowledge that the difference between the T$_{\mathrm{eff}}$ of the stars might lead to marginal differences in the position of the continuous and therefore contributing to the systematic errors. Nevertheless, it is unlikely that this effect could cause a T$_{c}$ trend as the one observed.

\subsection{Stellar Activity} \label{sec.activity}
Analyzing 211 Sun-like stars, \citet{Spina2020} showed that stellar activity can modify the EWs of absorption lines in stellar spectra and hence the fundamental parameters and chemical abundances. However, \citet{Spina2020} found no evidence that the chemical  pattern produced by planet engulfment events could be caused instead by different activity levels of  members of the same stellar association (see their Figure 6). Despite these results, we measured the activity index $\log R'_{HK}$ of both components of WASP-160 to compare their activity levels. 

Since our GRACES spectra do not include the Ca \textsc{ii} H and K lines, we used the infrared triplet lines of ionized calcium (Ca \textsc{ii} -IRT) at 8498, 8542 and 8662 {\AA}. The excess flux in these lines is well correlated to the standard  activity indicator $\log R'_{HK}$ \citep{Busa2007, Martin2017}. After obtaining the absolute excess fluxes of the Ca \textsc{ii} - IRT lines following \citet{Lorenzo2016}, we converted these values to the $\log  R'_{HK}$-index using Eq. 10 in \cite{Martin2017}. We found that both stars present very similar activity indices: $\langle \log R'_{HK}\rangle=-4.714$ $\pm$ 0.0434 and $-4.766$ $\pm$ 0.0434 for WASP-160A and WASP-160B, respectively\footnote{These values place  both stars near the limit between \textit{active} an \textit{inactive} stars \citep{Henry1996} and more active than the Sun \citep{Egeland2017}.}. Thus, we found no evidence that could link the observed abundance differences between the WASP-160 stars to their chromospheric activity status. These results agree with our analysis of the TESS light curves, where we did not find evidence of flares or rotational modulation in WASP-160A nor WASP-160B.




\begin{figure*}[ht!]
\centering
\includegraphics[width=.45\textwidth]{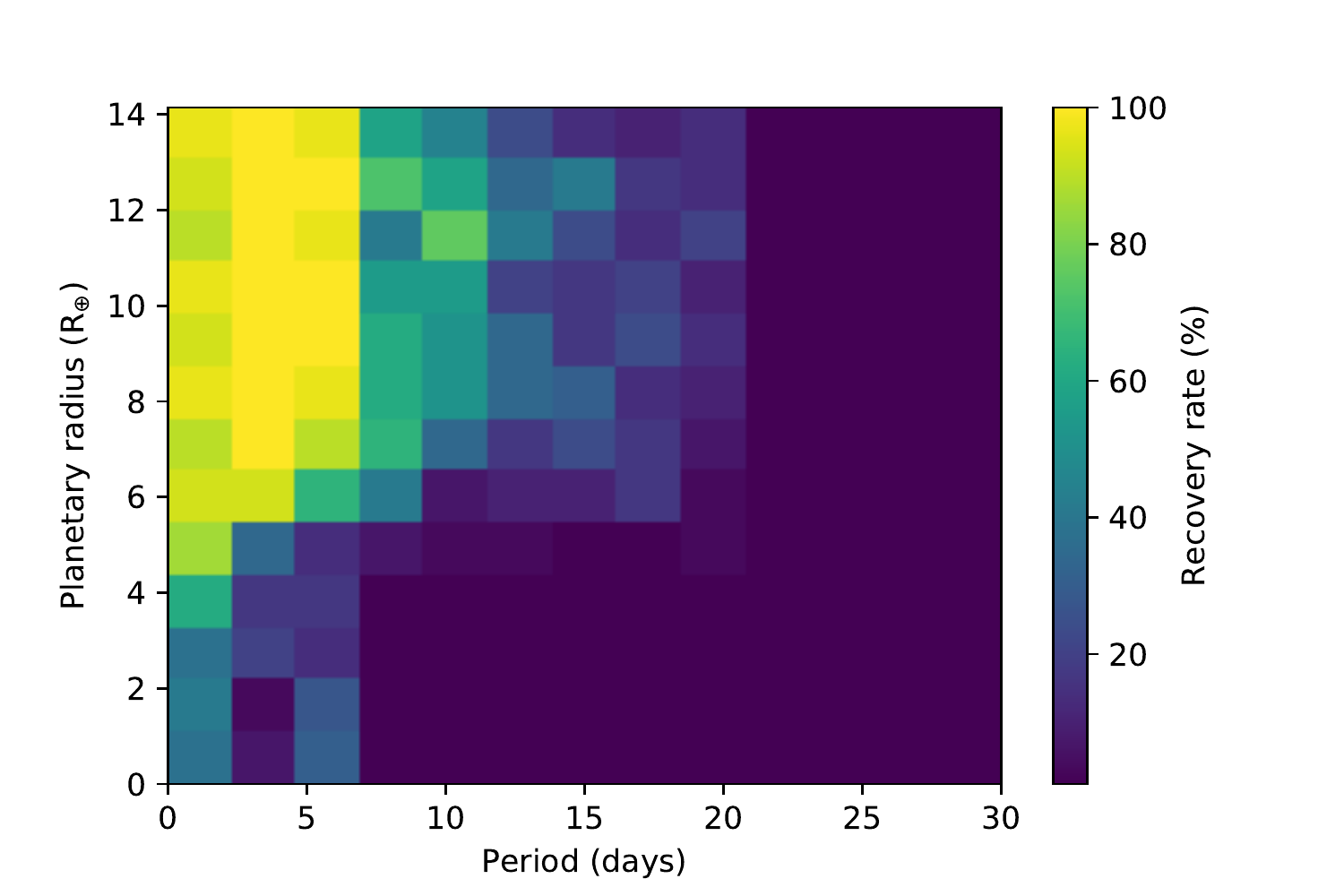}
\includegraphics[width=.45\textwidth]{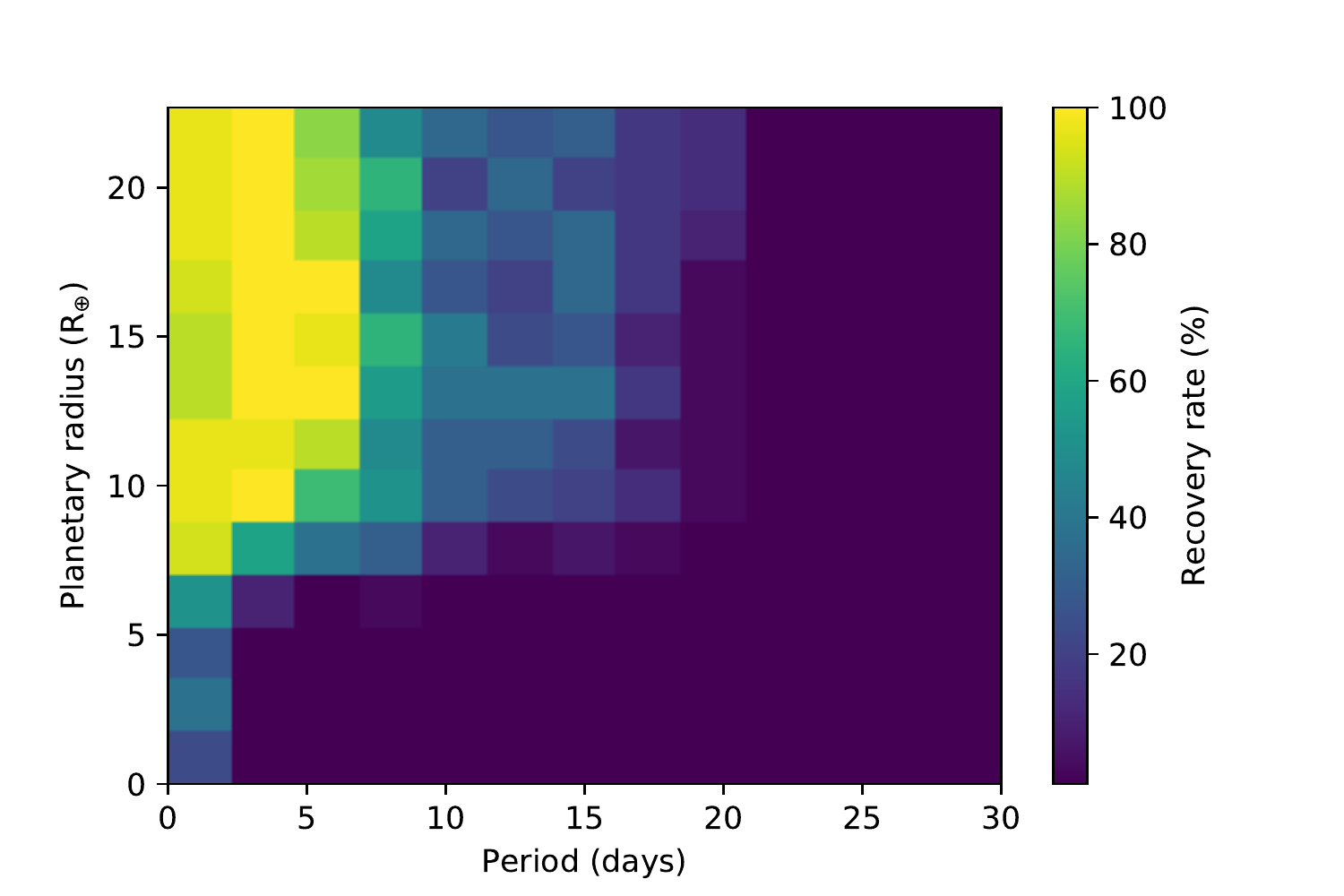}
\caption{Injection-recovery tests performed on the TESS light curves of WASP-160A (left) and WASP-160B (right). Light and dark colors indicate high  and low recovery rates, respectively. Planetary radii are corrected by the contamination due to the companion. \label{fig.injrec}}
\end{figure*}

\section{Planetary Analysis}\label{sec.planetary}


\subsection{Search for Additional Transiting Planets Around WASP-160A and WASP-160B from TESS Data} \label{search.tess}

In order to constrain possible scenarios that could account for the chemical pattern observed in Figure \ref{fig.Tcond.todo}, we searched for signs of additional transiting planets around both stars of the system using the TESS photometric data presented in Section {\ref{tess}}. For the B component, we employed the light curve obtained in Section {\ref{tess}} but with the data-points of all the five transits of the planet removed. For the A star, we followed the same procedure applied for the secondary component to get the light curve. We picked up the pixel with the highest flux centered on the position of WASP-160A and performed single aperture photometry on it (Figure \ref{fig.w160AB.tess}). As a consequence of the flux contamination by WASP-160B, this light curve showed some residual from the transits around the companion that we eliminated by removing the data-points at the times of all the transits. Lastly, we employed a Savitzky-Golay filter to smooth some remanent non-corrected systematics. 

Once we obtained the detrended light curves of both components, we ran on each one separately the Transit Least Squares code \citep[TLS,][]{Hippke2019A}, an algorithm designed to search for periodic transits from time-series photometry. TLS did not detect any feature that could be clearly attributed to the transit of a new planet around any of the components.


Nonetheless, we performed two injection-recovery tests to evaluate the detection limits of transiting planets in the TESS light curves of both stars in the system. We explored the planetary radius-orbital period parameter-space, $R_{P}$-$P$, from 0.0 to 12.0 R$_{\earth}$\footnote{R$_{\earth}$ is the radius of Earth.} with a step of 1.0 R$_{\earth}$ for the planetary radius and from 0 to 30.0 days with a step of 2.5 days for the orbital period. The transit signals injected into the TESS light curves were generated with the \texttt{BATMAN} code \citep{Kreidberg2015}  assuming circular orbits and equatorial transits (inclination of 90 degrees). We considered a positive detection when the period recovered by the TLS code is within 5$\%$ of any half-multiple of the injected period. In Figure {\ref{fig.injrec}}, we present the results of these tests. Planetary radii in the Y-axes are already corrected by a factor that accounts for the contamination due to the companion. In the case of the B component, this value was computed by \texttt{EXOFASTv2} through the parameter FITDILUTE, whilst for the primary star we used the factor estimated for the companion and the dilution parameter ``D'', as defined in \citet{Sullivan2015}:

\begin{equation}{\label{eq1}}
    D = \frac{F_A + F_B}{F_A}
\end{equation}

\noindent where F$_A$ and F$_B$ are the fluxes of WASP-160A and WASP-160B, respectively, integrated in the TESS band. 

The most important conclusion that arises from Figure {\ref{fig.injrec}} is that we might be able to detect, with high probability ($>$ 80$\%$), transiting planets larger than $\sim$ 5.5 R$_{\earth}$ and $\sim$ 7 R$_{\earth}$ with orbital periods shorter than $\sim$ 7 days around WASP-160A and WASP-160B, respectively. We do not expect a detection of planets at far distances from the stars (periods $>$ 20 days) or for small planets (radii $<$ 5.5 R$_{\earth}$) with orbital periods longer than $\sim$ 7 days in the case of WASP-160A, and small planets (radii $<$ 7 R$_{\earth}$) with orbital periods longer than $\sim$ 2.5 days for WASP-160B. The detection probabilities range from 20 to 80$\%$ for all the planets that occupied the parameter-space in between.

 \begin{figure*}[ht!]
\centering
\includegraphics[width=.45\textwidth]{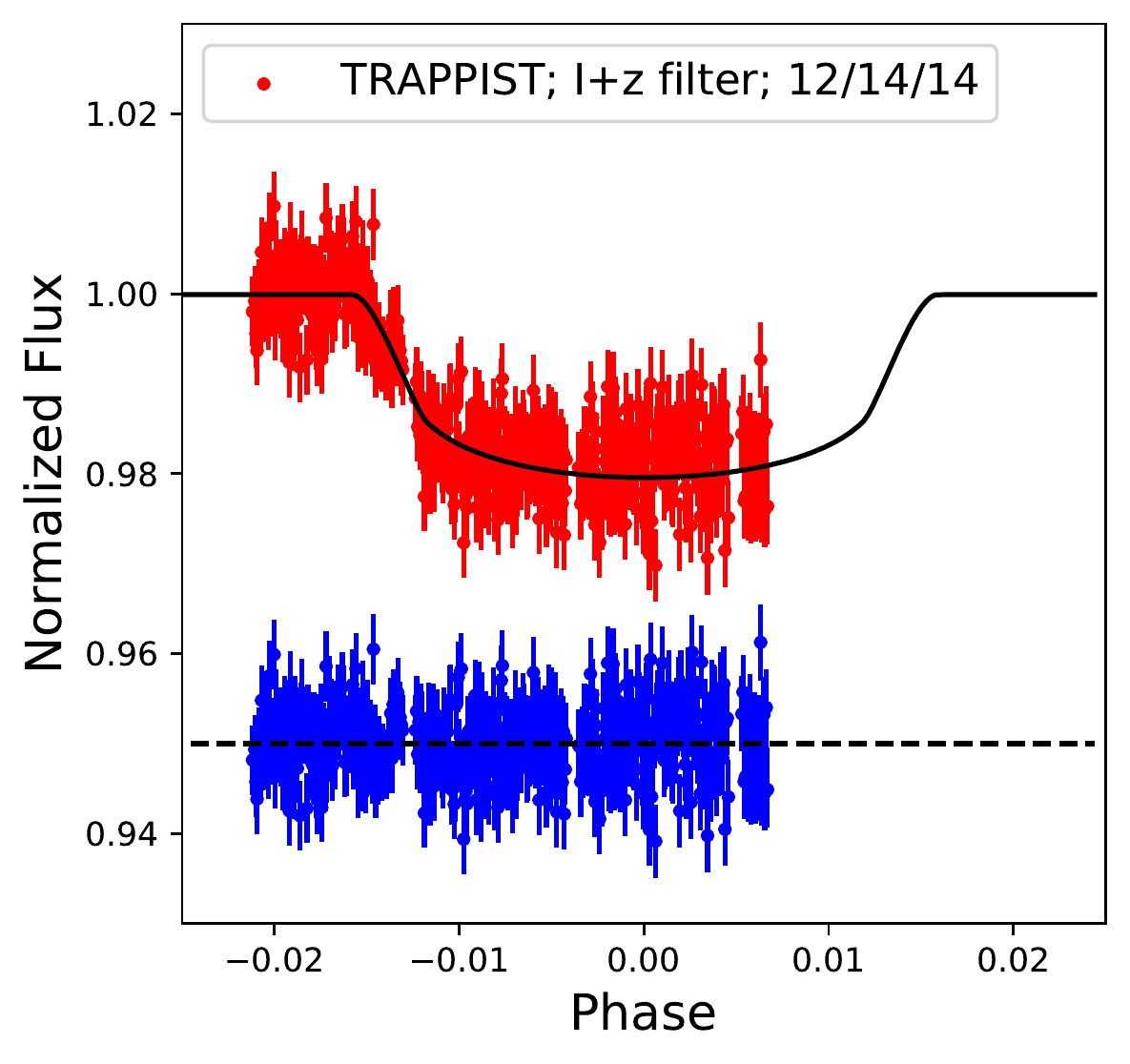}
\includegraphics[width=.45\textwidth]{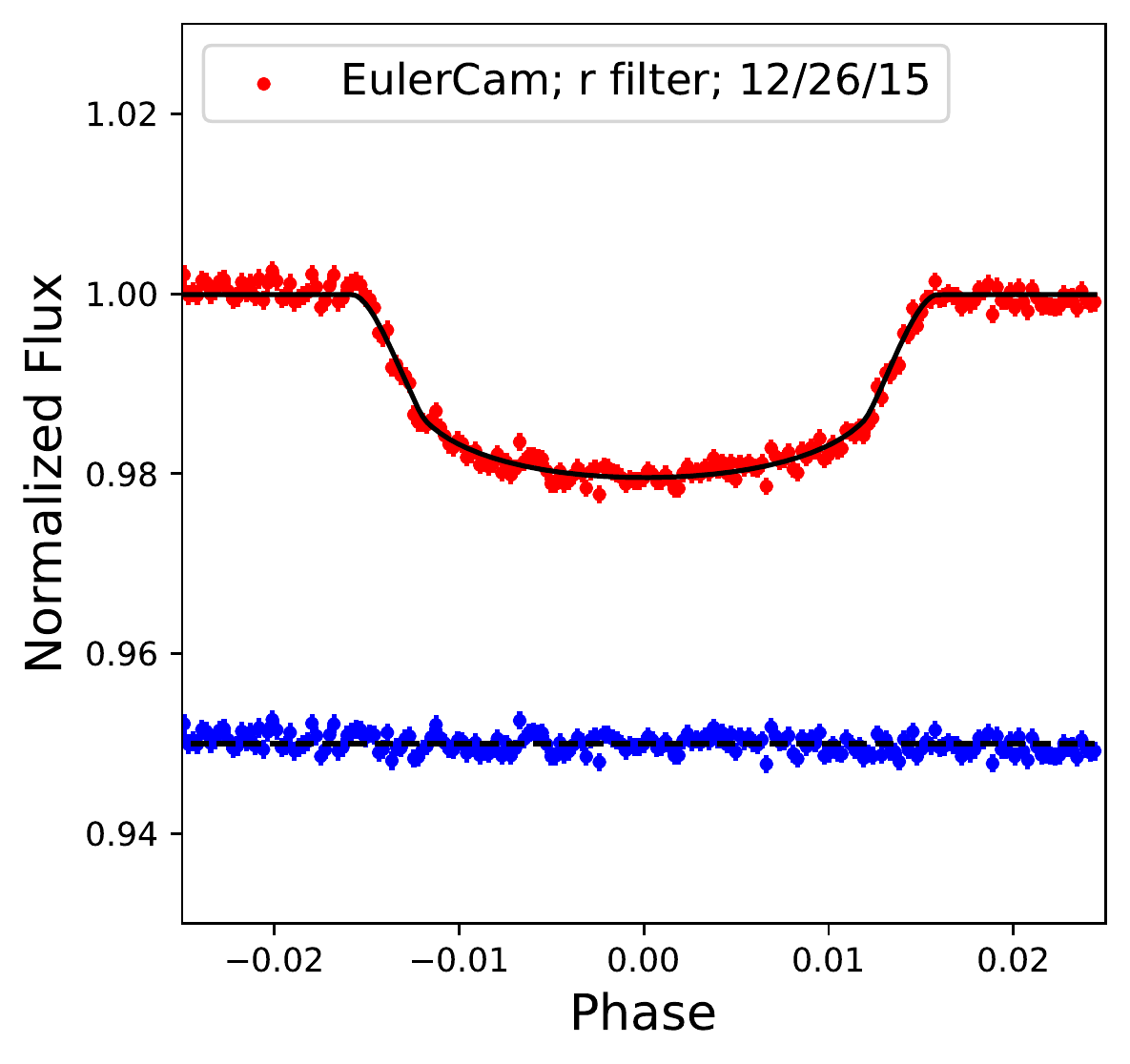}
\includegraphics[width=.45\textwidth]{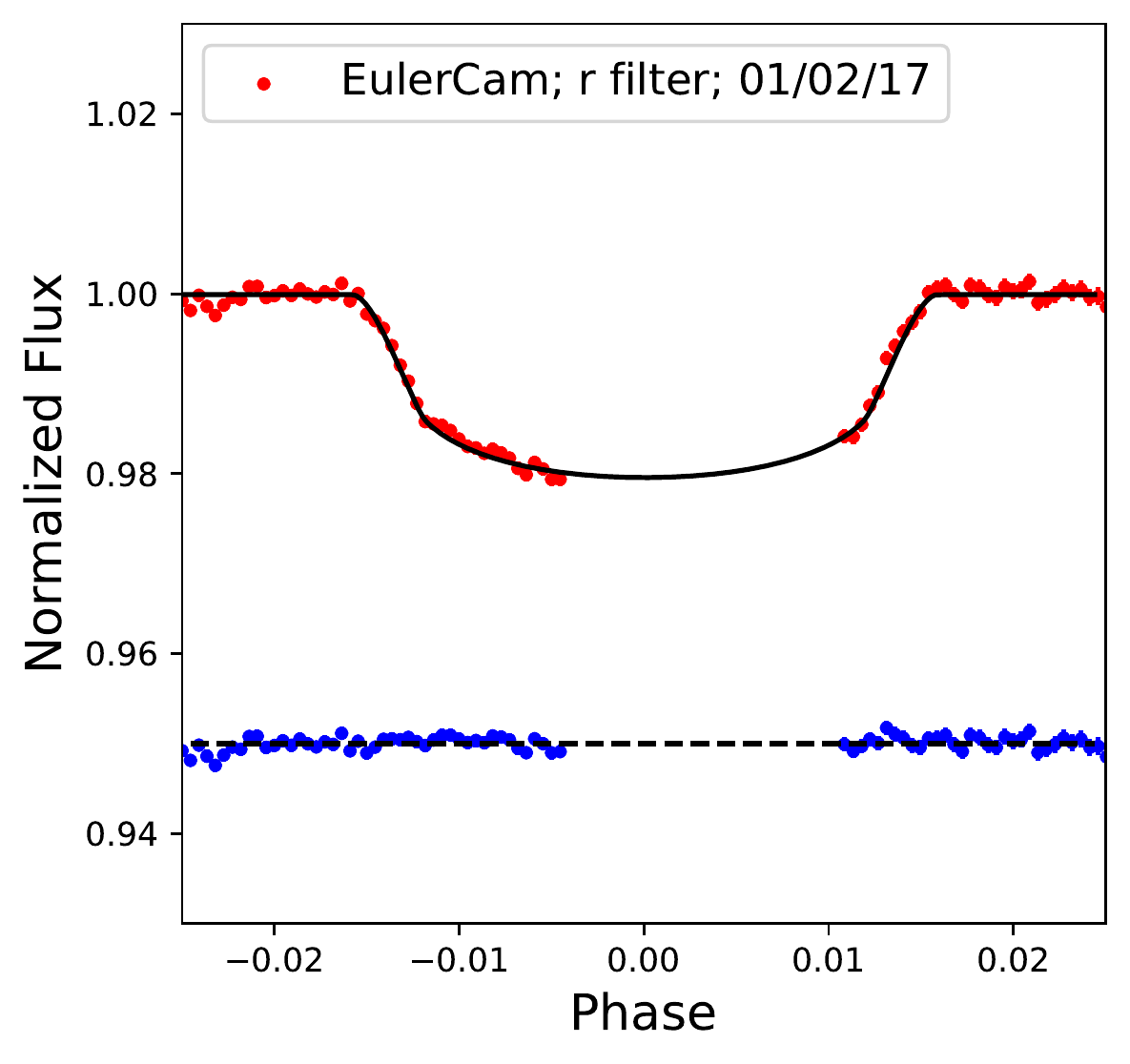}
\includegraphics[width=.45\textwidth]{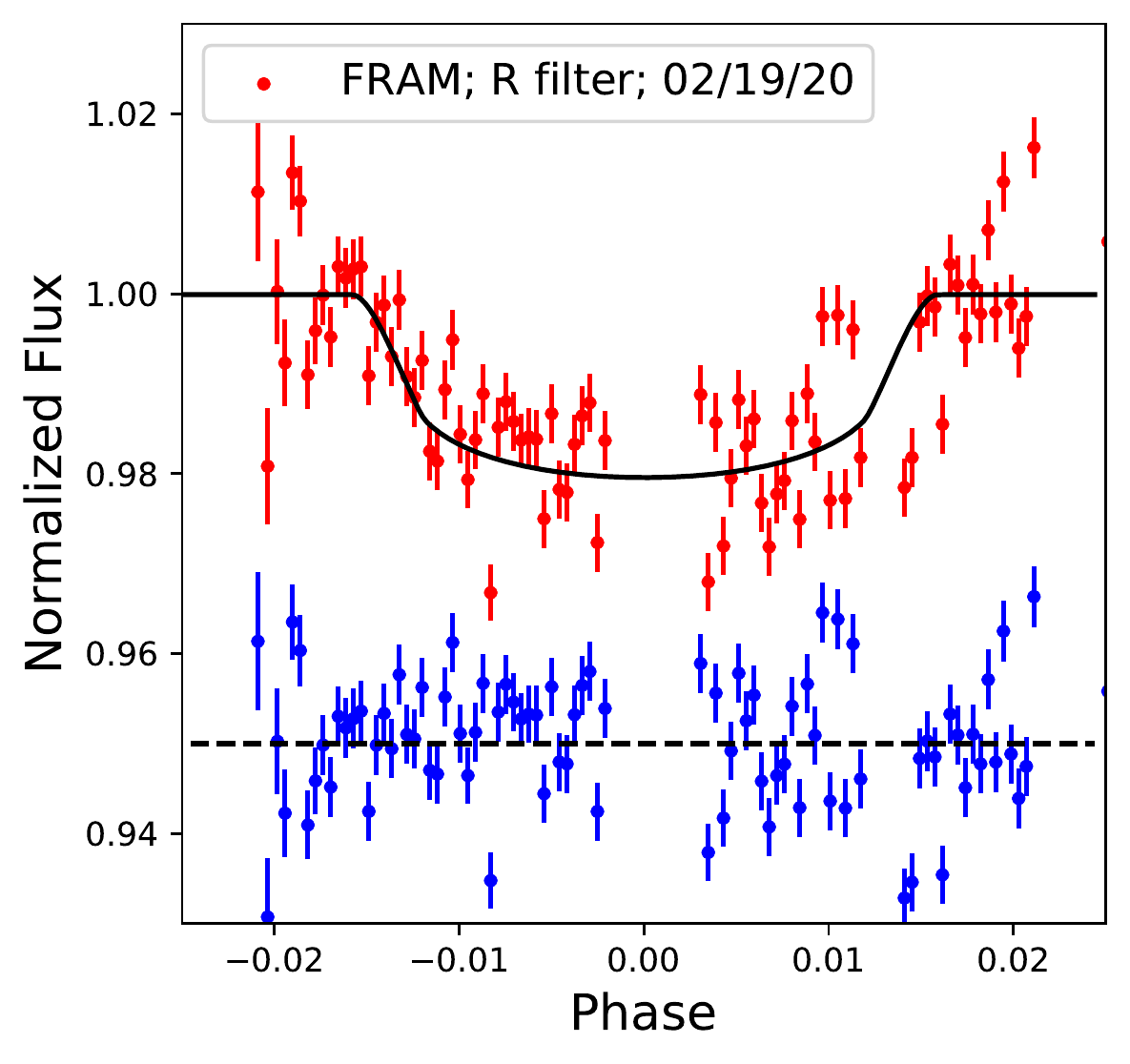}
\caption{Ground-based light curves of WASP-160B. In the top part of each panel, the black solid line is the best-fit model from \texttt{EXOFASTv2}. The residual (data minus model) are shown at the bottom of each panel (in blue), where a horizontal dashed black line is plotted for reference. The observations are phase-folded to the period of the transiting planet. The names of the instruments and observation dates
are denoted in boxes in the upper part of each panel.
\label{fig.groundLCRV}}
\end{figure*}

\begin{figure}[ht!]
\centering
\includegraphics[width=.45\textwidth]{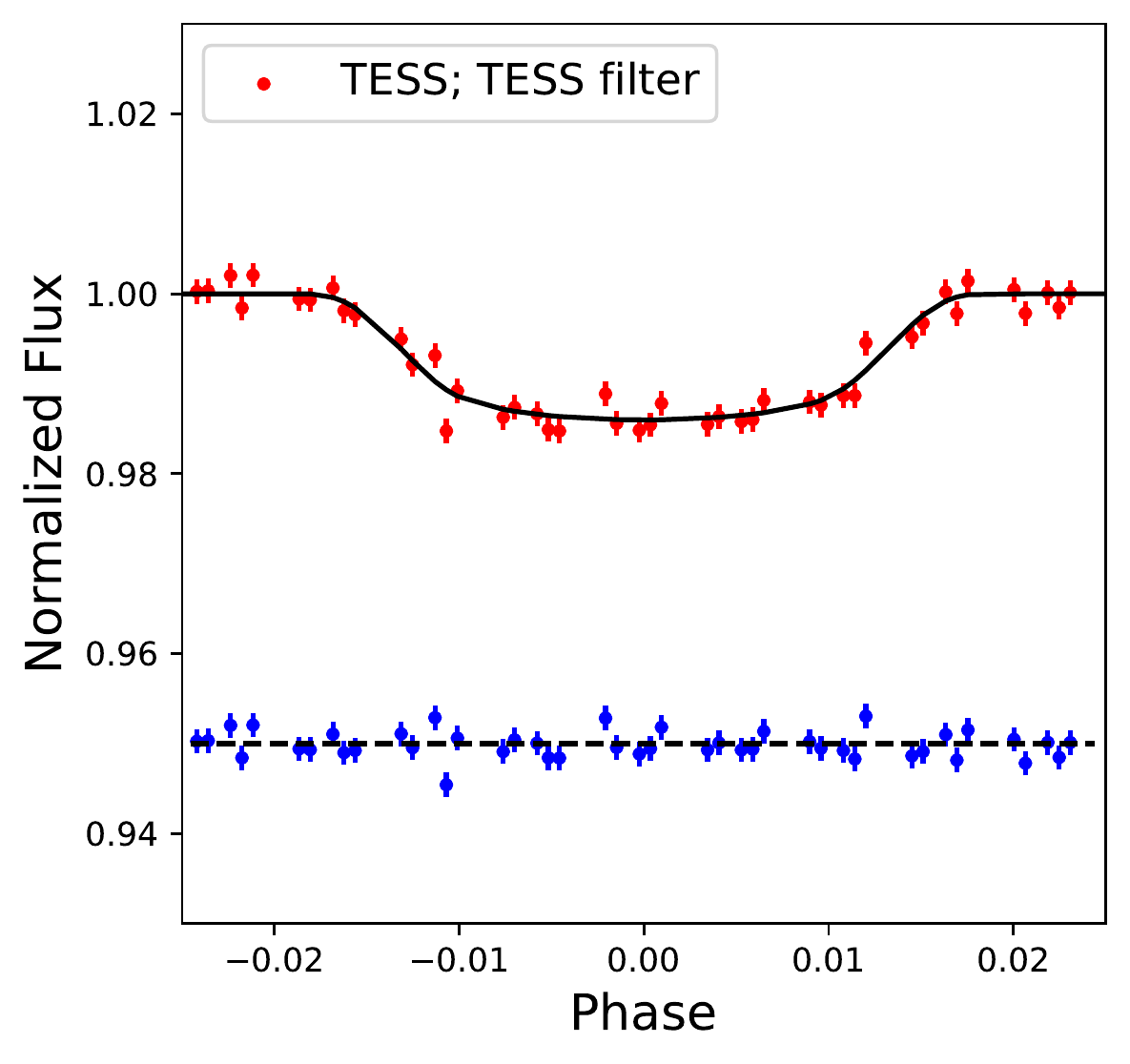}
\caption{Phase folded TESS light curve (red) of WASP-160B. The black solid line is the best-fit model from \texttt{EXOFASTv2}. The residuals are plotted below (in blue), where a horizontal dashed black line is plotted for reference. \label{fig.tessLC}}
\end{figure}

\begin{figure}[ht!]
\centering
\includegraphics[width=.45\textwidth]{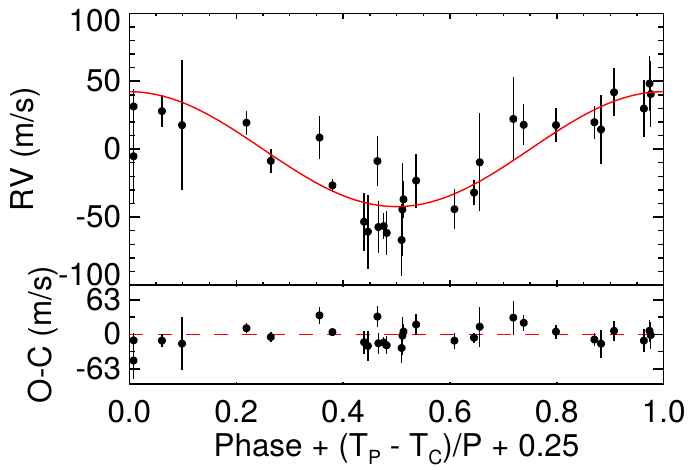}
\caption{Radial velocity measurements of WASP-160B b from CORALIE phase folded to the best determined period from our global model. The red solid line is the best-fit model from \texttt{EXOFASTv2}. The residuals are plotted below. Here $T_{P}$ is the time of periastron, $T_{C}$ is the time of conjunction or transit, and $P$ is the orbital period.} \label{fig.RV}
\end{figure}

\subsection{Refined Planet Properties of WASP-160B b}{\label{exofastv2}}

We derived planetary and stellar properties of WASP-160B with the latest version of the \texttt{EXOFASTv2} code \citep{Eastman2013, Eastman2019}. We performed  a simultaneous fit through a differential evolution Markov Chain Monte Carlo to the transit light curves and RV measurements presented in Section~\ref{sec.data}, the broadband photometry, \textit{Gaia} DR2 parallax for WASP-160B and A$_{\rm V}$ extinction, and the effective temperature and metallicity derived in Section~\ref{sec.atmospheric.parameters}. All stellar properties are summarized in Table~\ref{table.all.stellar.parameters}. 

We set Gaussian priors for the parameters with independent constraints, namely  T$_{\mathrm{eff}}$, [Fe/H], parallax, and extinction, with widths corresponding to their one-sigma uncertainties. As for WASP-160A, we set an upper limit to the extinction of 0.04 to avoid unrealistic solutions based on the galactic dust maps by \citet{Schlafly2011}. Additionally, we used a Gaussian prior for the orbital period with its uncertainty derived by LE19, because we do not include the WASP light curve in this analysis which contributes significantly to the determination of the orbital period.  We assumed a circular orbit. 
 We used as starting points in 
 M$_{\mathrm{\star}}$, R$_{\mathrm{\star}}$, and age, the values derived with \texttt{q$^{2}$} presented in Section~\ref{sec.other.parameters}. Since these three parameters are not independent from the \texttt{EXOFASTv2} analysis, we do not restrict them or set them as Gaussian priors in the fitting. 
 
 The physical properties of the planetary system are then derived by simultaneously fitting 
 all the available data described above to a keplerian orbital motion, the transit model \citep{mandelagol2002},
 a single-star SED, and a set of stellar models, either MIST \citep{Choi2016} or YY \citep{Demarque2004}. In the case of the TESS light curve, we accounted for the 30-min cadence of the observations by setting EXPTIME to 30 and NINTERP to 10, and for the dilution in the light curve due to WASP-160A, by fitting for the FITDILUTE parameter.  
 Thus, we obtain two best-fit solutions for WASP-160B, one where the stellar properties are constrained with the MIST stellar models and a second one that is constrained with the YY stellar models.
 Each of the best-fit solutions are attained by starting the MCMC chains several times until all the data sets were optimally fitted. 
 We ran \texttt{EXOFASTv2} several times for each solution, using the same Gaussian priors, and after the first run, the starting points for the previous MIST or YY run, respectively. Each MCMC ran until the two convergence criteria were reached \citep[number of independent draws T$_z   >$ 1000 and Gelman-Rubin statistic $<$ 1.01; as described in][]{Eastman2013}. 
 As a check on our assumption that the planetary orbit is circular, we ran \texttt{EXOFASTv2} allowing the eccentricity to be a free parameter using the same priors and the YY best-fit solution as a starting point. It resulted in a solution with a small eccentricity (0.03 $^{+0.03}_{-0.021}$),  with a $\sim$97\% probability of being spurious \citep[see][]{Lucy1971}; thus confirming our choice for a circular orbit. 
 

 The derived physical properties for WASP-160B for the YY best-fit solution are presented in Table~\ref{table.all.stellar.parameters} for the host star, and in Table~\ref{table.planetary.parameters} of the Appendix \ref{appendixA}, for the transiting planet.  We adopt the median values of the posterior distributions of each best-fit solution, and 68$\%$ confidence intervals as their one-sigma uncertainties.
 Both YY and MIST best-fit solutions are equivalent and consistent within their uncertainties. We have a slight preference for the best-fit solution derived from the YY models, because it allows us to compare our stellar properties with our other results minimizing systematic uncertainties arising from the choice of stellar models.   
 Our refined planetary parameters are consistent within the errors with those found by LE19, but are more precise because of the refined stellar properties. In Figures \ref{fig.groundLCRV} and \ref{fig.tessLC}, we show the resulting  best-fit models for the ground-based and the TESS transit light curves, respectively, and in Figure \ref{fig.RV} we show the best-fit model to the radial velocity measurements. 

Additionally, we performed a linear fit to the mid-transit times computed with \texttt{EXOFASTv2}. Given that the resulting O-C values are all within 1$\sigma$ of a linear ephemeris, we do not find evidence of transit timing variations (TTVs) in the system. However, it is important to notice that the small number of transits analyzed in this work prevent us from fully discarding the existence of additional planets around WASP-160B with the TTVs analysis. Certainly, an intensive photometric follow-up to significantly increase the number of complete and high-quality transits available would place more robust constraints on the presence of intermediate-mass short-period planets in the system.

\section{Discussion}\label{sec.discussion}

\subsection{Origin of the chemical differences}
WASP-160A and B show a non-negligible difference in their chemical compositions (Figure \ref{fig.abun.vs.Z}) which also seems to correlate with the condensation temperature (Figure \ref{fig.Tcond.todo}). In particular, we observe that WASP-160B is slightly depleted in volatiles ($-$0.035 dex, on average) and enhanced in refractory elements (+0.033 dex, on average) when compared to WASP-160A. To date, this kind of curious chemical pattern has been previously reported only in the binary system WASP-94 \citep{Teske2016a}, where each component hosts a hot Jupiter planet.  

As discussed in Section \ref{sec.differences.abundances}, the $\Delta$[X/H]$_{B-A}$ versus T$_{c}$ correlation that we have detected cannot be only due to our particular choice of stellar parameters. The T$_{c}$ trend is still present when we recomputed the abundances using: \textit{(i)} spectroscopic fundamental parameters with solar reference, \textit{(ii)} photometric T$_{\mathrm{eff}}$ and trigonometric $\log g$ values, and \textit{(iii)} the fundamental parameters from LE19. Moreover, given the good agreement between our adopted $\Delta$T$_{\mathrm{eff}}$ value and those derived from independent techniques, it is unlikely that our computed temperatures could have produced the observed pattern. Also, we found no evidence that could link the observed abundance differences with their measured chromospheric activity. 
Under the reasonable and common assumption that the components of binary systems share their initial composition \citep[e.g.,][]{Desidera2006, Andrews2019, Hawkins2020, Nelson2021}, since they form coevally from the same molecular cloud, it is unlikely that their dissimilar abundances are caused by different Galactic chemical evolution \citep[e.g.,][]{Adibekyan2014, Nissen2015} or different stellar environment effects \citep[e.g.,][]{Onehag2014}. Alternatively, as discussed in several previous studies of binary stars with similar components, the elemental abundance differences might be connected to processes of planetary formation and evolution \citep[e.g.,][]{Ramirez2011, Ramirez2015, Teske2016a, Teske2016b, Saffe2015, Saffe2017, Oh2018, Nagar2020}. 

What is the origin of the differences in the abundance of volatile elements? It has been suggested that the formation of a giant planet with large volatile-rich envelopes would cause a decrease in the overall metallicity of its host star, and therefore also producing a  depletion in the abundances of volatile elements \citep[e.g.,][]{Ramirez2011, Ramirez2015, Tucci2014, Teske2016a, Teske2016b}. Hence, the observed depletion of volatiles in WASP-160B, relative to WASP-160A, could be explained by the formation of the transiting giant planet, WASP-160B b, and/or other giant planets not yet detected in the system. As a first-order approximation, we can explore if the mass of WASP-160B b could account for the observed volatile differences using the corrected formula from \citet{Ramirez2011}:

\begin{equation}{\label{eq2}}
\Delta[M/H] = \log \left[ \frac{(Z/X)_{cz} M_{cz} + (Z/X)_{p} M_{p} } {(Z/X)_{cz} (M_{cz}+M_{p})} \right]
.
\end{equation}

From the tracks by \citet{Siess2000}, using the mass of WASP-160B, we estimated a present-day convective envelope mass $M_{cz}$ of 0.047 M$_{\odot}$. 
We adopt 0.1 for the planet metallicity $(Z/X)_{p}$ based on the solar system giant planets  \citep[e.g.,][]{Guillot2005, Fortney2010, Ramirez2011}, 
 0.28 M$_{J}$ for the mass of WASP-160B b (see Table \ref{table.planetary.parameters}), and 0.019 for  the stellar convection zone metallicity $(Z/X)_{cz}$ estimated by scaling the solar value of $(Z/X)_{\odot}$ of 0.0134 \citep{Asplund2009} to the metallicity of WASP-160B ([Fe/H] =  0.151 dex). Interestingly, under these assumptions, the expected depletion in WASP-160B $\Delta[M/H]$ is $\sim$0.0104 dex, which is close to the observed difference in volatiles. Even a closer expected value of $\sim$0.025 dex can be obtained if the planet metallicity is slightly larger\footnote{$(Z/X)_{p}$ could vary in the $\sim$0.03–0.5 range \citep{Thorngren2016}.}. 

We note that the $M_{cz}$ that appears in Equation \ref{eq2} corresponds to the convective zone at the time of planet formation, which for gas giants is suggested to occur on timescales shorter than $\sim$10 Myr\footnote{Although see, e.g., \citet{Pfalzner2014} } \citep[e.g.,][]{Pollack1996, Guilera2011}. Alternative nonstandard models of stars with severe episodic accretion computed by \citet{Baraffe2010} predict the $M_{cz}$ of a 1 M$_{\odot}$ star can reach its present-day thin size earlier than $\sim$10 Myr. Thus, as discussed in several previous studies \citep[e.g.,][]{Chambers2010, Ramirez2011, Ramirez2014, Onehag2011, Melendez2016, Teske2016a, Nissen2018, Spina2016, Spina2018, Tucci2019}, the hypothesis that planet formation (like WASP-160 B b) could have chemically altered the surface abundance pattern (i.e., depletion of volatiles) of its parent star would be plausible. However, as mentioned in \citet{Ramirez2014}, the results from \citet{Baraffe2010} also imply that the particular episodic accretion history of a star that forms planets will determine whether the chemical signature is imprinted or not. Moreover, recent stellar accreting models \citep{Kunitomo2018} show that the evolution of the convective zone is relatively close to that of standard (non-accreting) pre-main-sequence evolutionary tracks for a $\sim$1 M$_{\odot}$ star\footnote{For a $\sim$1 M$_{\odot}$ star the $M_{cz}$ reaches a minimum only after $\sim$20-30 Myr \citep{Kunitomo2018}.}. 


Standard stellar models suggest that solar analog stars begin fully convective and reach their present-day low $M_{cz}$ only after $\sim$30 Myr \citep{DAntona1994, Murray2001, Serenelli2011}. In this case, if instead we assume $M_{cz}$ = 0.4 M$_{\odot}$ for a $\sim$0.9 M$_{\odot}$ star at $\sim$10 Myr according to the standard model of \citet{Serenelli2011}, the expected depletion in WASP-160B to explain the planet mass is only $\sim$0.002 dex, which is well below the observed offset in volatiles. If instead we assume a larger $M_{cz}$ of 0.7 M$_{\odot}$ (at $\sim$5 Myr), we expect a lower offset of 0.0008 dex.

Alternatively, we can reverse the Equation \ref{eq2} and determine the mass of material that could explain the missing $\sim$0.035 dex of volatiles in WASP-160B. Considering the standard early convective envelopes of $M_{cz}$ = 0.4 M$_{\odot}$ and $M_{cz}$ = 0.7 M$_{\odot}$ for $(Z/X)_{p}$ = 0.1, we found that the mass required to explain the offset in volatiles is about 6 $M_{J}$ and 13 $M_{J}$, respectively. Therefore, in these cases, the formation of one or more gas-giant planets around WASP-160B, additional to the one already detected, would be needed to explain the observed depletion of volatiles.

The RV data from LE19 do not suggest the presence of additional giant planets in WASP-160B. However, we note that both the precision and time coverage of the CORALIE observations are not enough to fully reject the presence of a giant long period planet and/or additional lower mass bodies. As detailed in Section \ref{search.tess}, we found no signs of additional transiting planets using the available TESS photometry, although we were only able to discard, with a probability higher than 80$\%$, the presence of large, close-in planets ($P$ $\lesssim$ 7 days) with $R_P$ $\gtrsim$ 7 R$_{\oplus}$ (see Figure \ref{fig.injrec}). However, as in the case of the RV data from LE19, the analyzed TESS photometric data is inconclusive with regards to the presence of smaller-size and/or longer-period planets.  Furthermore, if additional bodies actually formed around WASP-160B, the non-detection of these planets might suggest that could have been ejected from the system by planet-planet interactions. 

On the other hand, we need to combine the hypothesis above for the origin of the \textit{missing} volatiles with possible scenarios to explain the enrichment of refractory elements in WASP-160B, relative to WASP-160A, and the weak upward T$_{c}$ trend which is also observed in Figure \ref{fig.Tcond.todo}. The chemical differences in refractories could be interpreted as the signature of either abundance depletion in WASP-160A or enhancement in WASP-160B: 

-i) In the first case, the difference in refractory elements can be considered within the scenario proposed by M09 and \citet{Ramirez2011}: the formation of rocky material (e.g., terrestrial planets, planetesimals) might result in a depletion of refractory elements (in comparison to volatiles) in the convective envelope of the host star, relative to its initial composition. Under this framework, we might interpret that refractory elements were sequestered from the protostellar nebula of WASP-160A to form rocky bodies while the volatiles were accreted onto the star. Using the model outlined in \citet{Chambers2010}, via the \texttt{terra} code\footnote{\url{https://github.com/ramstojh/terra}} \citep{Galarza2016}, we find that $\sim$4.5 $M_{\oplus}$, where $M_{\oplus}$ is the mass of Earth, of material (with an Earth-like composition) are needed to explain the observed offset in the refractory elements which, as discussed in Sec. \ref{sec.differences.abundances}, also seem to show a weak correlation with T$_{c}$. As before, here we have assumed a present-day convection zone mass $M_{cz}$ = 0.039 M$_{\odot}$ for WASP-160A. If, as discussed before, instead we use larger convective envelopes at younger stellar ages such as $M_{cz}$ = 0.2 M$_{\odot}$ (at 15 Myr) and $M_{cz}$ = 0.4  $M_{\odot}$ (at 10 Myr), then $\sim$21 $M_{\oplus}$ and $\sim$47 $M_{\oplus}$ would be required to explain the observed trend\footnote{Very similar values are obtained using the $M_{cz}$ computed from the models of \citet{Pinsonneault2001} and \citet{Murray2001}.} \footnote{For all the $M_{cz}$ values analyzed here, the best solution adopted from \texttt{terra} is reached when the value of the $\chi_{r} ^{2}$, which results from comparing the observed differential abundances with those computed by \texttt{terra}, is minimum. In all cases, we have explored the complete space parameter given by the mass of meteoritic-like and earth-like material between 0 and 60 $M_{\oplus}$.}. 

To date, no planets have been reported to orbit WASP-160A. LE19 obtained eight CORALIE spectra of WASP-160A and found no evidence of any large-amplitude RV variability. In particular, they found that the RV measurements are stable within $\sim$40 ms$^{-1}$. Thus, it is unlikely that this star harbors a short-period giant planet similar to that found around WASP-160B. However, both the precision and time coverage of the RV data are not enough to fully reject the presence of a giant long period planet and/or a low mass body. 

In an effort to investigate the existence of additional planets around WASP-160A which might account for the chemical pattern, as is detailed in Section \ref{search.tess}, we searched for transiting planets using the available TESS photometry. We found no transit signals and we were able to discard, with a probability higher than 80$\%$, the presence of large, close-in planets ($P$ $\lesssim$ 7 days) with radii larger than 5.5 R$_{\oplus}$ (see Figure \ref{fig.injrec}). However, as in the case of the RV data from LE19, the analyzed TESS photometric data is inconclusive with regards to the presence of small and/or longer-period planets ($P$ $\geq$ 20 days). Long-term RV follow-up and/or additional photometric monitoring of  WASP-160A might help to constrain the possible formation of planets around this star. For example, the TESS extended mission will observe WASP-160 in sectors 32 and 33 broadening the search to planets with periods within a few tens of days and increasing the chance of monotransit detections \citep[e.g.,][]{Cooke2019}. Future space telescopes within the next decade will also extend the search for planets, with observations of reflected light with Roman Space Telescope \citep[e.g.,][]{Carrion2020}, and to smaller planets and long periods, if WASP-160 is within the selected fields of PLATO \citep{Rauer2014}. 

We also tried to provide initial restrictions on the formation of planetesimals or/and asteroids around WASP-160A by investigating the presence of circumstellar material around this star. \citet{Saffe2016} proposed that the refractory-elements depleted in $\zeta^{2}$ Ret, relative to its twin binary companion $\zeta^{1}$ Ret, are locked-up in the rocky material that produce the debris disk observed around this object  \citep[although see][]{Faramaz2018}. Since the computed missing mass of refractories in WASP-160A is roughly compatible with the masses estimated for debris disks around solar-type stars \citep[3-50 M$_{\oplus}$;][]{Krivov2008}, we could consider a similar scenario to explain the lack of refractories in WASP-160A. Nevertheless, we find no signs of infrared excess on the SEDs, obtained in Sections \ref{check.Teff} and \ref{sec.other.parameters}, that could be associated with circumstellar dust around WASP-160A. We caution, however, that the magnitudes employed to obtain the SEDs are restricted to mid-IR bands only and therefore the analysis is inconclusive with regards to the presence of cold debris disks at large radial distances. Additional observations in the far-IR and (sub-)millimeter wavelength regions and/or high-contrast imaging combined with high-precision polarimetry, would provide better constraints on the formation of planetesimals around WASP-160A\footnote{Also, the SED of WASP-160B does not show signs of IR excess.}.

-ii) In the second scenario, after the initial decrease in the overall metallicity and volatile abundances in WASP-160B, allegedly caused by the formation of the Saturn-mass planet and possibly additional ones, late-time accretion of refractory-rich planetary material onto WASP-160B could have enhanced the abundance of high-T$_{c}$ elements producing the observed trend. The rocky bodies could have been pushed or dragged onto WASP-160B during the inward migration of the gas giant planet \citep[e.g.,][]{Pinsonneault2001, Ford2008, Sandquist2002, Ida2008, Raymond2011, Mustill2015} or by perturbations caused by its wide stellar companion \citep[e.g.,][]{Kaib2013}. Again, using \texttt{terra} we can estimate the amount of refractory-rich material accreted by WASP-160B for different convective envelopes sizes to reproduce the observed T$_{c}$ trend. For the present-day convective envelope of 0.047 M$_{\odot}$, we find that $\sim$6 M$_{\oplus}$ of material (with an Earth-like composition) are needed to explain the observed T$_{c}$ trend in the refractory elements.  This value rises up to $\sim$24 $M_{\oplus}$ if we consider $M_{cz}$ = 0.2 $M_{\odot}$ (at 15 Myr) and up to $\sim$48 M$_{\oplus}$ for $M_{cz}$ = 0.4 M$_{\odot}$ (at 10 Myr). We caution that this result does not necessarily imply that WASP-160A did not engulf rocky material also, but instead would suggest that WASP-160B accreted a larger amount of material than the primary.

As suggested in the case of other binary systems \citep[e.g.,][]{Saffe2017, Oh2018}, it would be expected that the accretion of planetary material on the atmosphere of WASP-160B would also produce an enhancement in its abundance of lithium, relative to that of WASP-160A. Nevertheless, as we mention in Sec. \ref{sec.abundances}, for both stars in WASP-160 we only could derive an upper limit of their abundances (A(Li) $\lesssim$ 0.17 dex and A(Li) $\lesssim$ 0.6 dex for WASP-160A and B, respectively), which are normal for their old ages and T$_{\mathrm{eff}}$ values \citep{Ramirez2011, Aguilera2018}. Although these results might not support a pollution scenario, we should stress that the behavior of Li abundances in solar-type stars with planets is very complex, not yet fully understood, and still debated \citep[e.g.,][]{Chaboyer1998, Israelian2009, Pinsonneault2010, Ghezzi2010, Baumann2010, Ramirez2012, Figueira2014, Delgado2015, Mishenina2016}. Moreover, as discussed in \citet{Melendez2017}, it is difficult to quantify the net amount of Li enhancement because it depends on several factors such as the initial stellar lithium content at the time of accretion, the time of accretion, Li depletion since the planet accretion, and the real effect and efficiency of thermohaline mixing for the depletion of Li \citep[e.g.,][]{Theado2012}.

Finally, \citet{Booth2020} recently proposed an alternative hypothesis to that of M09. According to this new scenario, the depletion of refractories in the Sun does not arise from the sequestration of refractory material inside the planets themselves but instead from the gap, and pressure trap, created by the formation of a Jupiter analog planet that prevents the accretion of the dust (refractories) exterior to its orbit onto the star in contrast to the gas (volatiles). For the case of WASP-160, however, it is not clear how this mechanism would explain the chemical pattern observed in Figure \ref{fig.Tcond.todo}. If we consider only the planet orbiting WASP-160B, according to the hypothesis of \citet{Booth2020} this star should show low abundances of refractories relative to volatiles but we observe the opposite trend. On the other hand, if also a Jupiter analog formed around WASP-160A, in order to explain the chemical trend, the pressure trap created by this planet should have been more effective in impeding the accretion of refractory material onto the host star than the \textit{confirmed} Saturn-mass planet around WASP-160B. Although the latter could perhaps be related by the migration history of WASP-160B b or its mass, the observed dissimilar composition of volatiles is hard to reconcile since in this framework the abundances of these elements should not be affected by the gaps. A detailed simulation within the scenario of \citet{Booth2020} for WASP-160, which is beyond the scope of this paper, might provide additional constraints on how this mechanism could explain the observed chemical pattern for this binary system.

\subsection{Abundance differences and binary separation}
\citet[][hereafter RA19]{Ramirez2019} studied a sample of twelve binary systems (with and without planets) with available high-precision chemical abundances in the literature to search for correlations with stellar parameters and/or binary star characteristics. They found a weak correlation between the absolute value of elemental abundance differences and binary star projected separation. In particular, the differences seem to increase as the distance between the stars in each pair becomes larger. Moreover, the differences become more significant for species that are least abundant in the Sun's photosphere (e.g., Sr, Zr, Rb, Y, Ba) compared to those species that are more abundant (e.g., O, C; see their Fig. 9). RA19 proposed that this could be an effect produced by the chemical inhomogeneity of the gas clouds from which the binary stars formed. Furthermore, RA19 emphasized that their results do not compromise the significance of the T$_{c}$ trends previously found on binary systems and/or their interpretation as due to sequestering or ingestion of planetary material. Instead, the authors suggest that the effect of chemical inhomogeneity might blur the T$_{c}$ trends observed in some binaries. 

Using our computed abundances and the projected separation between the components of WASP-160, we checked if this system could also follow the weak correlation presented by RA19. We found that the WASP-160 data is consistent with the trend for elements with high solar absolute abundance (O, C), in which there is a small system-to-system $\Delta$[X/H] variation. The elements with a lower absolute abundance in the Sun tend to show larger relative differences, in comparison to the results obtained with O and C. Nevertheless, they are not as large as would be expected for a system with the projected separation of WASP-160. Overall, under the scenario proposed by RA19, the results for WASP-160 might suggest that the molecular cloud from which this pair formed was more chemical homogeneous relative to other analyzed systems. However, since the correlations presented by RA19 are of second order, an analysis with a larger sample of binary systems will allow us to better understand the results for WASP-160. 

\section{Summary and Conclusions}\label{sec.conclussions}

In this work, we have expanded the small sample of well-studied planet-hosting binary stars by performing a detailed characterization of the WASP-160 stellar pair. No planet has yet been reported around WASP-160A while WASP-160B hosts a short-period transiting Saturn-mass planet, WASP-160B b. For this planet, we also derived refined properties by performing a global fit that includes our precise stellar parameters, literature data, and new observations. Furthermore, based on TESS photometry, we constrained the presence of transiting planets around WASP-160A and additional ones around WASP-160B. Our injection-recovery analysis allowed us to discard, with high probability ($>$ 80$\%$), the presence of transiting planets with orbital periods $\lesssim$ 7~days, and radii $>$ 5.5~$R_{\mathrm{\oplus}}$ and $>$ 7~$R_{\mathrm{\oplus}}$ around WASP-160A and WASP-160B, respectively.

The stellar characterization included, for the first time, the computation of high-precision and strictly differential atmospheric parameters and chemical abundances of 25 elements based on high-quality spectra. The analysis revealed that WASP-160B is slightly but significantly depleted in volatiles and enhanced in refractory elements relative to WASP-160A. Therefore we found evidence of a nonzero slope trend between $\Delta$[X/H] and T$_{c}$. Interestingly, this curious chemical pattern showing both depletion of volatiles and enhancement of refractories has been previously reported only in the binary system WASP-94, where each star hosts a hot Jupiter planet \citep{Teske2016a}. 

We discussed that the depletion of volatiles could be explained by the formation of the observed Saturn-mass planet WASP-160B b if we assume the present-day $M_{cz}$ of the host star, but single or multiple additional planets (to add up to $\sim$6$-$13 $M_{J}$) are required when adopting a larger $M_{cz}$. On the other hand, the refractory-difference could be explained by \textit{(i)} the formation of rocky bodies (at least $\sim$4.5 $M_{\oplus}$) around WASP-160A or \textit{(ii)} the late accretion of at least $\sim$6 $M_{\oplus}$ of planet-like material by WASP-160B.

The formation of the confirmed short-period giant planet orbiting WASP-160B and later accretion of planet-like material by this star, likely triggered by the inward migration of the giant planet,  represents a tantalizing scenario to explain the observed anomalous chemical pattern. However, the detection limits provided by the current photometric and RV data along with the significant dependence on the assumed $M_{cz}$ prevent us from fully supporting this last hypothesis over the formation of additional (long period) giant planets in WASP-160B and those of the rocky kind around WASP-160A. Future long-term high-precision photometric and RV follow-up, as well as  high-contrast imaging observations, of WASP-160A and B, would provide further constraints and might reveal the real origin of the peculiar chemical differences observed.

\acknowledgments
This work is partially based on observations obtained at the Gemini Observatory, which is operated by the Association of Universities for Research in Astronomy, Inc., under a cooperative agreement with the NSF on behalf of the Gemini partnership: the National Science Foundation (United States), National Research Council (Canada), CONICYT (Chile), Ministerio de Ciencia, Tecnolog\'{i}a e Innovaci\'{o}n Productiva (Argentina), Minist\'{e}rio da Ci\^{e}ncia, Tecnologia e Inova\c{c}\~{a}o (Brazil), and Korea Astronomy and Space Science Institute (Republic of Korea). We are grateful to Dr. M. Lendl for kindly providing the transits presented in their study. We thank the anonymous referee for several comments and suggestions which helped us to improve the manuscript. This research has been partially supported by UNAM-PAPIIT IN-107518. E. J. and R. P. acknowledge DGAPA for their postdoctoral fellowships. E.M. acknowledges funding from the French National Research Agency (ANR) under contract number ANR-18-CE31-0019 (SPlaSH).

This study is based on observations obtained through the Gemini Remote Access to CFHT ESPaDOnS Spectrograph (GRACES). ESPaDOnS is located at the Canada-France-Hawaii Telescope (CFHT), which is operated by the National Research Council of Canada, the Institut National des Sciences de l’Univers of the Centre National de la Recherche Scientifique of France, and the University of Hawai’i. ESPaDOnS is a collaborative project funded by France (CNRS, MENESR, OMP, LATT), Canada (NSERC), CFHT and ESA. ESPaDOnS was remotely controlled from the international Gemini Observatory, a program of NSF’s NOIRLab, which is managed by the Association of Universities for Research in Astronomy (AURA) under a cooperative agreement with the National Science Foundation on behalf of the Gemini partnership: the National Science Foundation (United States), the National Research Council (Canada), Agencia Nacional de Investigaci\'on y Desarrollo (Chile), Ministerio de Ciencia, Tecnolog\'ia e Innovaci\'on (Argentina), Minist\'{e}rio da Ci\^{e}ncia, Tecnologia e Inova\c{c}\~{a}o (Brazil), and Korea Astronomy and Space Science Institute (Republic of Korea). 

This paper includes data collected by the TESS mission. Funding for the TESS mission is provided by the NASA Explorer Program. We would also like to thank the Pierre Auger Collaboration for the use of its facilities. The operation of the robotic telescope FRAM is supported by the grant of the Ministry of Education of the Czech Republic LM2018102. The data calibration and analysis related to the FRAM telescope is supported by the Ministry of Education of the Czech Republic MSMT-CR LTT18004 and MSMT/EU funds CZ.02.1.01/0.0/0.0/16-13/0001402. This study has made use of SIMBAD database operated at CDS, Strasbourg (France); and the NASA ADS database. This work presents results from the European Space Agency (ESA) space mission Gaia. Gaia data are being processed by the Gaia Data Processing and Analysis Consortium (DPAC). Funding for the DPAC is provided by national institutions, in particular the institutions participating in the Gaia MultiLateral Agreement (MLA).

%

\vspace{5mm}
\facilities{Gemini-North: 8.1 m (Gemini Remote Access to CFHT ESPaDOnS Spectrograph, GRACES), F/(Ph)otometric Robotic Atmospheric Monitor (FRAM), Transiting Exoplanet Survey Satellite (TESS)}


\software{ astropy \citep{Astropy}, 
Exofastv2 \citep{Eastman2013, Eastman2019}, 
q$^{2}$ \citep{Ramirez2015}, 
IRAF (distributed by the National Optical Astronomy Observatories, which are operated by the Association of Universities for Research in Astronomy, Inc., under cooperative agreement with the National Science Foundation), 
lightkurve \citep{Lightkurve}, 
BATMAN \citep{Kreidberg2015}, 
TLS \citep{Hippke2019A}, 
OPERA \citep{Martioli2012},
tpfplotter \citep{Aller2020},
terra \citep{Galarza2016}
       }




\appendix
\section{Additional tables} \label{appendixA}

 \begin{table*}[th!]
  \small
      \caption{Adopted Atomic Data and Measured Equivalent Widths}
         \label{table.lines}
     \centering
         \begin{tabular}{c c c c c c c}
            \hline\hline
Wavelength	&	Species	&	EP	&	$\log gf$	&	WASP-160A	&	WASP-160B	&	Sun	\\
({\AA})	&		&	(eV)	&		&	(m{\AA})	&	(m{\AA})	&	EW (m{\AA})	\\
\hline													
6587.61	&	6.0	&	8.537	&	$-$1.021	&	10.0	&	6.3	&	14.3	\\
5380.34	&	6.0	&	7.685	&	$-$1.570	&	14.4	&	9.7	&	20.8	\\
9078.28	&	6.0	&	7.480	&	$-$0.581	&	73.5	&	53.7	&	110.0	\\
9111.80	&	6.0	&	7.490	&	$-$0.297	&	97.5	&	74.0	&	140.0	\\
7468.27	&	7.0	&	10.340	&	$-$0.150	&	7.8	&	5.0	&	4.6	\\
8683.40	&	7.0	&	10.330	&	$-$0.050	&	10.0	&	6.3	&	8.0	\\
7775.39	&	8.0	&	9.146	&	0.002	&	33.5	&	24.2	&	44.5	\\
7774.16	&	8.0	&	9.150	&	0.220	&	43.0	&	31.5	&	58.7	\\
7771.94	&	8.0	&	9.146	&	0.352	&	49.5	&	33.5	&	69.2	\\

    \hline
         \end{tabular} 
\tablecomments{(This table is available in its entirety in machine-readable form.)}        
    
        \end{table*}

 \begin{table*}[h!]
  \small
      \caption{Condensation Temperatures, Number of Measured lines, and Elemental Abundances Relative to Solar ([X/H]) and Differential Between WASP-160A and WASP-160B ($\Delta$[X/H]$_{B-A}$)}
         \label{table.abundances}
     \centering
         \begin{tabular}{l c c c c c c c c}
            \hline\hline

Species	&	T$_{c}$ (K)	&	N$_{lines}$	&	[X/H] (dex)	&	Error (dex)	&	[X/H] (dex)	&	Error  (dex)	&	$\Delta$[X/H]$_{B-A}$ (dex)	&	Error (dex)	\\
	&		&		&	\multicolumn{2}{c}{(WASP-160A)}			&	\multicolumn{2}{c}{(WASP-160B)}			&		&		\\
\hline																	
C \textsc{i}	&	40	&	4	&	$-$0.021	&	0.094	&	$-$0.097	&	0.090	&	$-$0.049	&	0.005	\\
N \textsc{i}	&	131	&	2	&	0.560	&	0.064	&	0.538	&	0.068	&	$-$0.022	&	0.007	\\
O \textsc{i}	&	180	&	3	&	0.130	&	0.021	&	0.118	&	0.052	&	$-$0.010	&	0.023	\\
Na \textsc{i}	&	958	&	4	&	0.312	&	0.026	&	0.321	&	0.029	&	0.008	&	0.005	\\
Mg \textsc{i}	&	1336	&	3	&	0.141	&	0.079	&	0.172	&	0.073	&	0.031	&	0.006	\\
Al \textsc{i}	&	1653	&	3	&	0.263	&	0.059	&	0.294	&	0.060	&	0.031	&	0.007	\\
Si \textsc{i}	&	1310	&	28	&	0.165	&	0.062	&	0.181	&	0.074	&	0.015	&	0.007	\\
S \textsc{i}	&	664	&	3	&	0.259	&	0.137	&	0.243	&	0.154	&	$-$0.010	&	0.012	\\
K \textsc{i}	&	1006	&	1	&	$-$0.152	&	0.172	&	$-$0.200	&	0.167	&	$-$0.048	&	0.028	\\
Ca \textsc{i}	&	1517	&	8	&	0.161	&	0.051	&	0.170	&	0.052	&	0.017	&	0.008	\\
Sc \textsc{i}	&	1659	&	3	&	0.189	&	0.099	&	0.254	&	0.121	&	0.065	&	0.029	\\
Sc \textsc{ii}	&	1659	&	7	&	0.215	&	0.048	&	0.265	&	0.046	&	0.050	&	0.007	\\
Ti \textsc{i}	&	1582	&	18	&	0.197	&	0.055	&	0.252	&	0.054	&	0.055	&	0.006	\\
Ti \textsc{ii}	&	1582	&	2	&	0.150	&	0.004	&	0.222	&	0.012	&	0.069	&	0.010	\\
V \textsc{i}	&	1429	&	6	&	0.262	&	0.081	&	0.381	&	0.082	&	0.099	&	0.017	\\
Cr \textsc{i}	&	1296	&	10	&	0.149	&	0.057	&	0.202	&	0.058	&	0.053	&	0.006	\\
Cr \textsc{ii}	&	1296	&	3	&	0.149	&	0.012	&	0.211	&	0.006	&	0.067	&	0.012	\\
Mn \textsc{i}	&	1158	&	3	&	0.163	&	0.102	&	0.206	&	0.166	&	0.043	&	0.008	\\
Fe \textsc{i}	&	1334	&	84	&	0.137	&	0.051	&	0.145	&	0.052	&	0.008	&	0.005	\\
Fe \textsc{ii}	&	1334	&	10	&	0.151	&	0.079	&	0.164	&	0.080	&	0.012	&	0.007	\\
Co \textsc{i}	&	1352	&	5	&	0.161	&	0.065	&	0.171	&	0.057	&	0.070	&	0.007	\\
Ni \textsc{i}	&	1353	&	40	&	0.183	&	0.067	&	0.205	&	0.064	&	0.022	&	0.005	\\
Cu \textsc{i}	&	1037	&	3	&	0.353	&	0.179	&	0.410	&	0.205	&	0.056	&	0.019	\\
Zn \textsc{i}	&	726	&	1	&	$-$0.039	&	0.172	&	$-$0.075	&	0.166	&	$-$0.036	&	0.026	\\
Rb \textsc{i}	&	800	&	2	&	$-$0.391	&	0.172	&	$-$0.383	&	0.141	&	0.008	&	0.031	\\
Y \textsc{ii}	&	1659	&	2	&	0.163	&	0.024	&	0.223	&	0.032	&	0.060	&	0.010	\\
Zr \textsc{ii}	&	1741	&	1	&	0.041	&	0.172	&	0.067	&	0.166	&	0.026	&	0.028	\\
Ba \textsc{ii}	&	1455	&	2	&	0.063	&	0.047	&	0.102	&	0.032	&	0.039	&	0.018	\\
Eu \textsc{ii}	&	1356	&	1	&	0.305	&	0.172	&	0.388	&	0.166	&	0.083	&	0.026	\\

    \hline
         \end{tabular}

        \end{table*}

\begin{table*}[h!]
\caption{Median values and 68\% confidence interval for the Physical and Orbital Parameters of WASP-160B b}
\label{table.planetary.parameters}
\centering
\begin{tabular}{l c c c}
\hline\hline
Parameter	&	Units	&	Values YY (circular)	&	Values MIST (circular)	\\
\hline							
~~~~$P$\dotfill	&	Period (days)\dotfill	&	$3.7684935\pm0.0000016$	&	$3.7684935^{+0.0000016}_{-0.0000015}$	\\
~~~~$a$\dotfill	&	Semi-major axis (AU)\dotfill	&	$0.04541^{+0.00037}_{-0.00035}$	&	$0.04536^{+0.00067}_{-0.00056}$	\\
~~~~$R_P$\dotfill	&	Radius ($R_{\mathrm{J}}$)\dotfill	&	$1.093^{+0.022}_{-0.021}$	&	$1.093^{+0.023}_{-0.022}$	\\
~~~~$M_P$\dotfill	&	Mass ($M_{\mathrm{J}}$)\dotfill	&	$0.281\pm0.030$	&	$0.280\pm0.031$	\\
~~~~$\rho_P$\dotfill	&	Density (cgs)\dotfill	&	$0.266^{+0.033}_{-0.032}$	&	$0.266^{+0.033}_{-0.032}$	\\
~~~~$logg_P$\dotfill	&	Surface gravity (cgs) \dotfill	&	$2.765^{+0.048}_{-0.052}$	&	$2.764^{+0.048}_{-0.052}$	\\
~~~~$T_{eq}$\dotfill	&	Equilibrium temperature (K)\dotfill	&	$1099.5^{+9.0}_{-8.9}$	&	$1099.5^{+9.0}_{-8.9}$	\\
~~~~$\Theta$\dotfill	&	Safronov Number \dotfill	&	$0.0265\pm0.0029$	&	$0.0264\pm0.0029$	\\
~~~~$\langle F\rangle$\dotfill	&	Incident Flux (10$^{9}$ erg s$^{-1}$ cm$^{-2}$)\dotfill	&	$0.331^{+0.011}_{-0.010}$	&	$0.331\pm0.011$	\\
\smallskip\\\multicolumn{2}{l}{Primary transit parameters:}\smallskip\\							
~~~~$R_P/R_*$\dotfill	&	Radius of planet in stellar radii \dotfill	&	$0.12877^{+0.00093}_{-0.00095}$	&	$0.12878^{+0.00093}_{-0.00094}$	\\
~~~~$a/R_*$\dotfill	&	Semi-major axis in stellar radii \dotfill	&	$11.2^{+0.19}_{-0.18}$	&	$11.19^{+0.19}_{-0.18}$	\\
~~~~$i$\dotfill	&	Inclination (Degrees)\dotfill	&	$88.80^{+0.49}_{-0.33}$	&	$88.79^{+0.49}_{-0.33}$	\\
~~~~$b$\dotfill	&	Transit Impact parameter \dotfill	&	$0.234^{+0.061}_{-0.093}$	&	$0.235^{+0.061}_{-0.094}$	\\
~~~~$\delta$\dotfill	&	Transit depth (fraction)\dotfill	&	$0.01658\pm0.00024$	&	$0.01658\pm0.00024$	\\
~~~~$T_{FWHM}$\dotfill	&	FWHM transit duration (days) \dotfill	&	$0.10425^{+0.00047}_{-0.00046}$	&	$0.10424\pm0.00046$	\\
~~~~$\tau$\dotfill	&	Ingress/egress transit duration (days)	&	$0.01426^{+0.00057}_{-0.00058}$	&	$0.01426^{+0.00058}_{-0.00059}$	\\
~~~~$T_{14}$\dotfill	&	Total transit duration (days) \dotfill	&	$0.11852^{+0.00057}_{-0.00056}$	&	$0.11851^{+0.00057}_{-0.00056}$	\\
~~~~$P_T$\dotfill	&	A priori non-grazing transit prob \dotfill	&	$0.0778\pm0.0012$	&	$0.0778\pm0.0012$	\\
~~~~$P_{T,G}$\dotfill	&	A priori transit prob \dotfill	&	$0.1008\pm0.0017$	&	$0.1008\pm0.0017$	\\
~~~~$d/R_*$\dotfill	&	Separation at mid transit \dotfill	&	$11.20^{+0.19}_{-0.18}$	&	$11.19^{+0.19}_{-0.18}$	\\
~~~~$Depth$\dotfill	&	Flux decrement at mid transit \dotfill	&	$0.01658\pm0.00024$	&	$0.01658\pm0.00024$	\\
~~~~$u_{1R}$\dotfill	&	linear limb-darkening coeff \dotfill	&	$0.485^{+0.048}_{-0.049}$	&	$0.485\pm0.048$	\\
~~~~$u_{1r’}$\dotfill	&	linear limb-darkening coeff \dotfill	&	$0.503\pm0.025$	&	$0.504\pm0.025$	\\
~~~~$u_{1z’}$\dotfill	&	linear limb-darkening coeff \dotfill	&	$0.304^{+0.040}_{-0.041}$	&	$0.303\pm0.041$	\\
~~~~$u_{1TESS}$\dotfill	&	linear limb-darkening coeff \dotfill	&	$0.400\pm0.046$	&	$0.401\pm0.046$	\\
~~~~$u_{2R}$\dotfill	&	quadratic limb-darkening coeff \dotfill	&	$0.199\pm0.049$	&	$0.198^{+0.050}_{-0.049}$	\\
~~~~$u_{2r’}$\dotfill	&	quadratic limb-darkening coeff \dotfill	&	$0.189\pm0.033$	&	$0.189\pm0.033$	\\
~~~~$u_{2z’}$\dotfill	&	quadratic limb-darkening coeff \dotfill	&	$0.231\pm0.047$	&	$0.231\pm0.047$	\\
~~~~$u_{2TESS}$\dotfill	&	quadratic limb-darkening coeff \dotfill	&	$0.222\pm0.049$	&	$0.221\pm0.048$	\\
~~~~$A_D$ (TESS)\dotfill	&	Dilution from neighboring stars \dotfill	&	$0.281\pm0.019$	&	$0.281\pm0.019$	\\
\smallskip\\\multicolumn{2}{l}{Radial velocity parameters:}\smallskip\\							
~~~~$T_C$\dotfill	&	Time of conjunction (BJD$_{\mathrm{TDB}}$)\dotfill	&	$2457383.65491\pm0.00012$	&	$2457383.65491\pm0.00012$	\\
~~~~$T_0$\dotfill	&	Optimal conjunction Time (BJD$_{\mathrm{TDB}}$)\dotfill	&	$2457504.24670\pm0.00011$	&	$2457504.24670^{+0.00011}_{-0.00010}$	\\
~~~~$T_P$\dotfill	&	Time of Periastron (BJD$_{\mathrm{TDB}}$)\dotfill	&	$2457383.65491\pm0.00012$	&	$2457383.65491\pm0.00012$	\\
~~~~$T_S$\dotfill	&	Time of eclipse (BJD$_{\mathrm{TDB}}$)\dotfill	&	$2457385.53915\pm0.00012$	&	$2457385.53915\pm0.00012$	\\
~~~~$T_A$\dotfill	&	Time of Ascending Node (BJD$_{\mathrm{TDB}}$) \dotfill	&	$2457386.48128\pm0.00012$	&	$2457386.48128\pm0.00012$	\\
~~~~$T_D$\dotfill	&	Time of Descending Node (BJD$_{\mathrm{TDB}}$)\dotfill	&	$2457384.59703\pm0.00012$	&	$2457384.59703\pm0.00012$	\\
~~~~$K$\dotfill	&	RV semi-amplitude (m s$^{-1}$) \dotfill	&	$39.9^{+4.2}_{-4.3}$	&	$39.8^{+4.2}_{-4.3}$	\\
~~~~$M_P\sin i$\dotfill	&	Minimum mass (M$_{\mathrm{p}}$)\dotfill	&	$0.281\pm0.030$	&	$0.280\pm0.031$	\\
~~~~$M_P/M_*$\dotfill	&	Mass ratio \dotfill	&	$0.000305\pm0.000033$	&	$0.000304\pm0.000033$	\\
~~~~$\gamma_{\rm CORALIE}$	&	Relative RV Offset (m s$^{-1}$)\dotfill	&	$-6141.7^{+3.2}_{-3.3}$	&	$-6141.7^{+3.2}_{-3.3}$	\\

\hline																																		
\end{tabular}

\end{table*}


\newpage

\bibliography{Jofre-2020}{}
\bibliographystyle{aasjournal}



\end{document}